\documentclass[reprint,amsmath,amssymb,aps,longbibliography]{revtex4-2}
\usepackage{amsmath}
\usepackage{graphicx}
\graphicspath{ {./} }

\usepackage{xcolor}
\usepackage{upgreek}

\usepackage{dcolumn}
\usepackage{bm}
\usepackage{color}
\usepackage{mathtools}
\usepackage[normalem]{ulem}
\usepackage[english]{babel}
\usepackage{xr-hyper}
\usepackage[colorlinks=true,linkcolor=blue,urlcolor=blue,citecolor=blue]{hyperref}

\usepackage{upgreek}
\usepackage{graphicx}
\usepackage{dcolumn}
\usepackage{bm}
\usepackage{color}
\usepackage{mathtools}
\usepackage[normalem]{ulem}
\usepackage[english]{babel}
\usepackage{xr-hyper}
\usepackage[colorlinks=true,linkcolor=blue,urlcolor=blue,citecolor=blue]{hyperref}

\newcommand{\del}[1]{\sout{#1}}  
\renewcommand{\del}[1]{}  

\newcommand{\Cite}[1]{Ref.~\cite{#1}}
\newcommand{\fig}[1]{Fig.~\ref{#1}}

\newcommand{\Fig}[1]{Figure~\ref{#1}}
\newcommand{\eq}[1]{Eq.~(\ref{#1})}

\newcommand{\eqs}[2]{Eqs.~(\ref{#1}) and (\ref{#2})}

\newcommand{\App}[1]{Appendix~\ref{#1}}

\newcommand{\roundbk}[1]{\left({#1}\right)}
\newcommand{\squarebk}[1]{\left[{#1}\right]}

\newcommand{\av}[1]{\vert{#1}\rvert}

\begin{document}

\title{Towards Relating Fragile-To-Strong Transition to Fragile Glass}

\author{Chin-Yuan Ong$^1$}
\author{Chun-Shing Lee$^1$}
\author{Xin-Yuan Gao$^1$}
\author{Qiang Zhai$^2$}
\author{Rui Shi$^3$}
\author{Hai-Yao Deng$^4$}
\email[Email: ]{dengh4@cardiff.ac.uk}

\author{Chi-Hang Lam$^1$}
\email[Email: ]{C.H.Lam@polyu.edu.hk}
\address{$^1$Department of Applied Physics, Hong Kong Polytechnic University, Hong Kong, China \\
$^2$School of Physics, MOE Key Laboratory for Nonequilibrium Synthesis and Modulation of Condensed Matter, Xi’an Jiaotong University, Xi’an 710049, China \\
$^3$Zhejiang Province Key Laboratory of Quantum Technology and Device, School of Physics,
Zhejiang University, Zheda Road 38, Hangzhou 310027, China \\
$^4$School of Physics and Astronomy, Cardiff University, 5 The Parade, Cardiff CF24 3AA, Wales, UK
}
	
\date{\today}

\begin{abstract}
Glass formers are in general classified as strong or fragile depending on whether their relaxation rates follow Arrhenius or super-Arrhenius temperature dependence. There are however notable exceptions such as water, which exhibit a fragile-to-strong (FTS) transition and behave as fragile and strong respectively at high and low temperatures. In this work, the FTS transition is studied using a distinguishable-particle lattice model previously demonstrated to be capable of simulating both strong and fragile glasses [Phys. Rev. Lett. 125, 265703 (2020)]. Starting with a bimodal pair-interaction distribution appropriate for fragile glasses, we show that by narrowing down the energy dispersion in the low-energy component of the distribution, a FTS transition is observed. The transition occurs at a temperature at which the stretching exponent of the relaxation is minimized, in agreement with previous molecular dynamics simulations. 
\end{abstract}

\maketitle
	
	
\section{Introduction}
\label{sec:intro}
	Glasses are produced when liquids are supercooled below the glass transition temperature \(T_{g}\)  \cite{stillinger2013review,biroli2013review,arceri2020}. Depending on how dramatically the relaxation rate of a glass changes with respect to the temperature upon cooling, it is then classified as a strong glass or a fragile glass. For strong glass, the dynamics follows Arrhenius law. In contrast, it shows a super-Arrhenius temperature dependence for fragile glass.
	However, there exist many types of liquids that do not obey this classification, such as water~\cite{ito1999,angell1993}, silica~\cite{saksaengwijit2004}, BeF\(_{2}\)~\cite{hemmati2001} and some metallic glasses ~\cite{evenson2012}. They are  fragile at high temperature, but their dynamics obeys Arrhenius law characteristic of  strong glass at low temperature. This anomaly is known as fragile-to-strong (FTS) transition.

    Water is the most widely studied material exhibiting a FTS transition. Several hypotheses based on mode-coupling theory (MCT)~\cite{gallo2000}   and Adam-Gibbs theory~\cite{ito1999} were proposed to account for the phenomenon. 
    Subsequently, Stanley and co-workers explained the dynamic abnormalities with a crossover from a high-density liquid to a low-density one at a line which is known as Widom line~\cite{xu2005}. It is defined as the maxima of thermodynamics response functions that emanate from a proposed liquid-liquid critical point.
 
    Alternatively, Tanaka proposed an explanation based on a two-state model by describing the FTS transition as a dynamic crossover from a high temperature state to a low temperature one ~\cite{tanaka2000,tanaka2003}. The existence of two distinct states were supported by experiments \cite{myneni2002,nilsson2010}.

	In this paper, we study the FTS transition using the distinguishable-particle lattice model (DPLM) of glass \cite{zhang2017}.
    The DPLM has recently been applied successfully to address Kovacs’ expansion gap paradox~\cite{lulli2020}, the connection between fragility and thermodynamics quantities~\cite{lee2020}, the heat capacity hysteresis in cooling-heating cycles with  large overshoots in fragile glasses~\cite{lee2021}, two-level systems at low temperature \cite{gao2022}, and a diffusion-coefficient power-law under a particle partial-swap algorithm \cite{gopinath2022}.
We will show that materials exhibiting a high fragility and a FTS transition are closely related and can differ only in the low-energy statistics of the particle pair interactions. Specifically, by simply altering the low-energy component of a pair-interaction energy distribution adopted in the DPLM, a fragile glass model can be turned into a system exhibiting a FTS transition.

	The DPLM is a lattice gas model defined on a two-dimensional (2D) square lattice with \textit{N} distinguishable particles \cite{zhang2017,lulli2020,lee2020,lee2021,gao2022,gopinath2022}. Periodic boundary conditions are assumed. At most one particle can occupy a site at a time. Voids of a density $\phi_{v}$ are introduced in the system as unoccupied sites. The total energy of the system is defined as 
	\begin{equation} 
		E = \sum_{<i,j>'}^{} V_{s_{i}s_{j}},
	\end{equation}
	where the sum is over nearest neighbor sites \textit{i} and \textit{j}, and the particle index $s_i = 1, \cdots, N$ represents which particle is located at site \textit{i}. The interaction energy $V_{kl}$ between particles \textit{k} and \textit{l} is randomly sampled from a pair interaction distribution \(g(V)\) before the simulation commences. 
The DPLM can produce both strong and fragile glasses by implementing a uniform-plus-delta bimodal interaction distribution given by \cite{lee2020}
	\begin{equation}
		g(V) = \frac{G_{0}}{\Delta V}+(1-G_{0})\delta(V-V_{1}),
		\label{gVfragile}
	\end{equation}
for $V_0\le V \le V_1$ with $\Delta V=V_1-V_0$. 
The thermodynamic parameter $G_0$ obeys ${0 < G_0 \le 1}$.
Strong and fragile glasses correspond to large and small $G_0$ respectively.
	
	The system follows void-induced dynamics governed by the Metropolis algorithm. Specifically, each particle can hop to an unoccupied nearest neighbor site at temperature  \(T\) at a rate
	\begin{equation}
		w  = w_{0}\exp{\squarebk{-\frac{E_{0}+\theta(\Delta E) ~ \Delta E}{k_{B}T}}}
		\label{w}
      \end{equation}
where \(\Delta E\) represents the change in total energy \textit{E} of the system after hopping, \(k_{B}=1\) is the Boltzmann constant, 
$\theta$ denotes the Heaviside  step function, \(E_{0}\) is an energy barrier offset, and \(w_{0}\) is a rate constant.

    \section{Fragile-to-strong transition}
	
	\label{pid}

	Dynamic behaviors of the DPLM largely depend on the pair interaction distribution \(g(V)\).
To enable a fragile-to-strong transition, the uniform-plus-delta distribution in \eq{gVfragile} is replaced in our main simulations in this work by a bi-delta distribution 
\begin{equation}
	g(V) = G_{1}\delta(V-V_{0})+(1-G_{1})\delta(V-V_{1}),
	\label{gVfts}
\end{equation}
where $V_0$ and $V_1$ ($V_0<V_1$) are the two possible interaction energies. The thermodynamic parameter $G_{1}$, analogous to $G_0$ in \eq{gVfragile}, represents the probabilistic weight of the lower interaction energy $V_0$. We take $V_0=0$ and $V_1=1$ so that  $\Delta V = V_1-V_0=1$.

\begin{figure}[tb]
\includegraphics[width=\columnwidth]{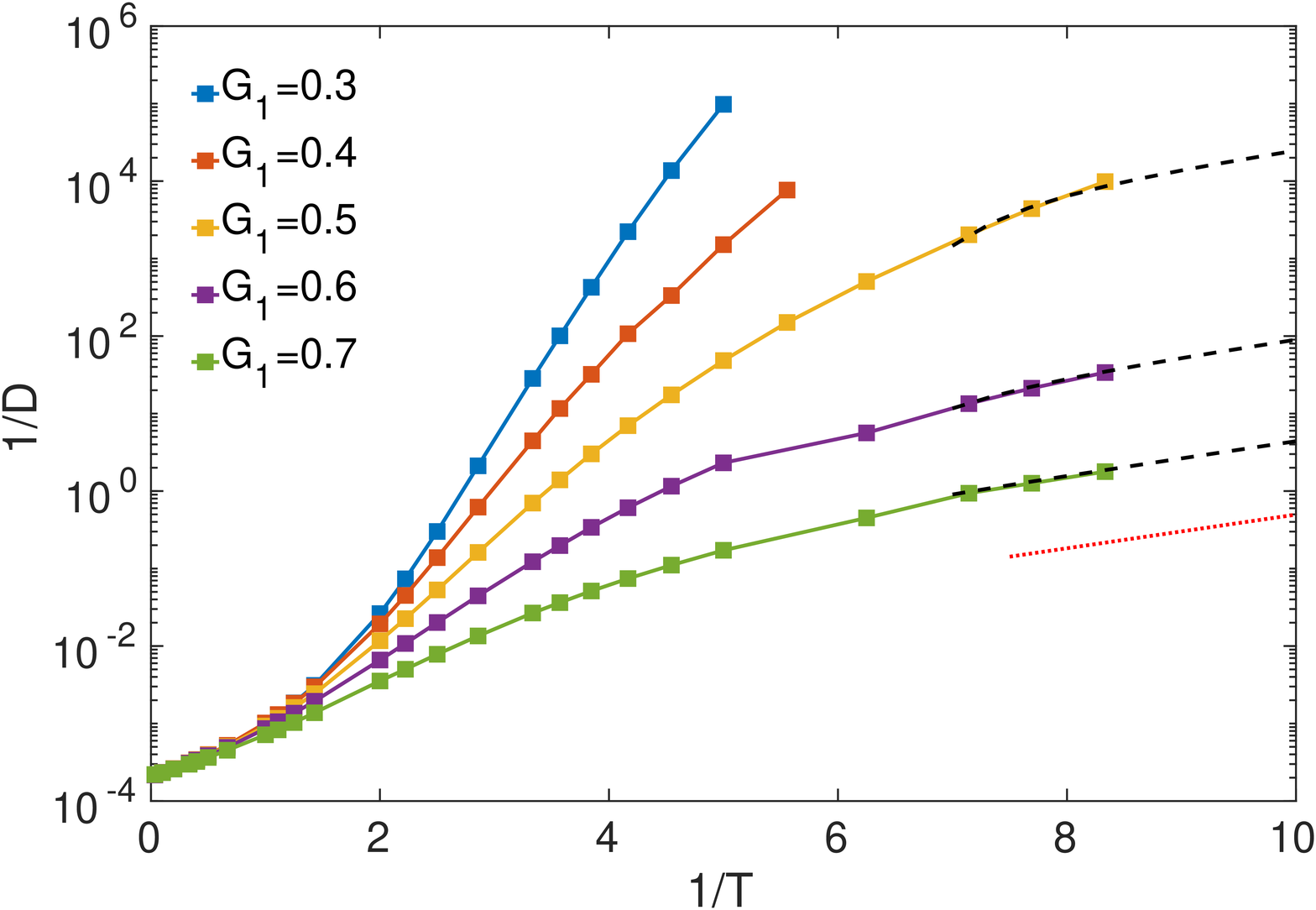}
\caption{Inverse particle diffusion coefficient \(1/D\) against inverse temperature \(1/T\) for  various probabilistic weight \(G_{1}\) of the low-energy state. We put \(E_{0}=0.5\) and $w_0=10^6$. The dashed lines represent fits to \eq{ArrheniusFit}. The dotted line following \eq{Arrhenius} is a guide to the eye.}
\label{vary_g0}
\end{figure}

    We have performed kinetic Monte Carlo simulations in 2D on a square lattice of length $L=100$ with the bi-delta pair interaction in \eq{gVfts}. We take $\phi_v=0.01$, $E_0=0.5$ and $w_0=10^6$. The diffusion coefficient $D$ at various temperature $T$ is measured (see \App{app:methods} for details). \Fig{vary_g0} shows an Arrhenius plot of $1/D$  for various values of $G_1$.
    For the smallest value of $G_1=0.3$ reported, we observe  for $1/T \alt 4$ a curvature in the semi-log plot, indicating a super-Arrhenius temperature dependence of $1/D$. This signifies a high-temperature fragile regime and the result is qualitatively similar to previous simulations in \Cite{lee2020} for moderately fragile glass with $G_0=0.3$ using the uniform-plus-delta distribution in \eq{gVfragile}. Nevertheless, a distinct feature  in \fig{vary_g0} is that for $1/T \agt 4$, the super-Arrhenius behavior turns into a sub-Arrhenius one, which was not observed in \Cite{lee2020}. The results are qualitatively similar for all other values of $G_1$ reported, although the super-Arrhenius nature diminishes at larger $G_1$. 

    The termination of the high-temperature super-Arrhenius behavior as temperature decreases marks a FTS transition. Far below the transition at very low temperature, the dynamics is Arrhenius as is typical of strong glass.
As lattice models are often more tractable analytically, the low-temperature Arrhenius behavior can be proven easily as follows.
For $k_BT \ll \Delta V$, excitation to interaction $V_1$ are suppressed so that only the lower-energy interaction $V_0$ is relevant. All possible particle hops are then constrained to those resulting at no energy change, i.e. $\Delta E = 0$. The rate $w$ of all energetically possible hops as given in \eq{w} then follow the same Arrhenius rate
\begin{equation}
w  = w_{0}\exp{\roundbk{-\frac{E_{0}}{k_{B}T}}}.
\label{wlowT}
\end{equation}
which defines the dynamic time scale. Simulations at different low temperatures are thus similar except for this trivial scaling factor in the hopping rate. We therefore arrive at the low-temperature Arrhenius form 
\begin{equation}
\label{Arrhenius}
D \sim \exp{\roundbk{ -\frac{E_0}{k_B T}}}, 
\end{equation}
The result signifies a strong behavior.

We observe from \fig{vary_g0} that the Arrhenius behavior in \eq{Arrhenius} is directly verified at low temperature for $G_1=0.7$ . For larger $G_1$, the observed trend is also consistent with a convergence to \eq{Arrhenius}. When approaching the  Arrhenius regime, we find that $D$ follows the empirical form
\begin{equation}
\label{ArrheniusFit}
D = D_0 \exp{\squarebk{ -\frac{E_0}{k_B T} + \roundbk{\frac{c_1}{T}-c_2}^{-4}}},
\end{equation}
where $D_0$, $c_1$ and  $c_2$ are constants and it reduces to \eq{Arrhenius} at low temperature. For example, for $G_1=0.5$, fitted values are $D_0=181$, $c_1=0.316$ and $c_2=1.30$.

\section{Modeling amorphous water}
\label{water}

We have demonstrated in the previous section that by adopting a different interaction distribution $g(V)$, a model of fragile glass can be turned into one exhibiting a FTS transition. We now further generalize it to model amorphous water more closely.     In water, both hydrogen and non-hydrogen bonds are present and they are crucial in studying its anomalies~\cite{titantah2013,debenedetti2003}. 
We envision that the two interaction energies $V_0$ and $V_1$ in \eq{gVfts} now describe hydrogen and non-hydrogen bonds respectively, an interpretation in line with a two-state picture of water~\cite{shi2018,shi2020}.

\begin{figure}[tb]
\includegraphics[width=\columnwidth]{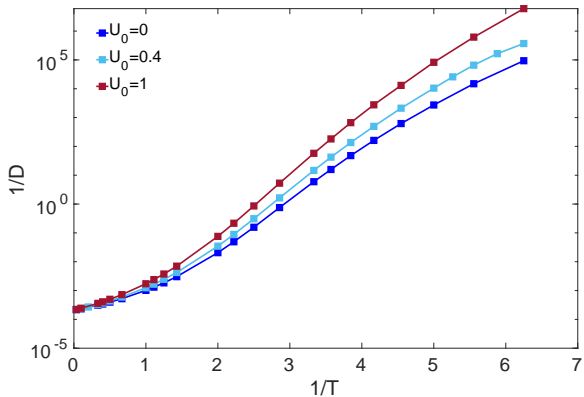}
\caption{Inverse particle diffusion coefficient \(1/D\) against inverse temperature \(1/T\) for  various $U_0$. We put \(G_{1}=0.38\), $E_0=0.5$ and $w_0=10^6$. }
\label{vary_U0}
\end{figure}

    We consider a probabilistic weight \(G_{1}=0.38\) of the low-energy interaction $V_0$. This value allows simulations to be performed within manageable run-time for observing both a high fragility at high temperature and a FTS transition at low temperature. 
\Fig{vary_U0} shows an Arrhenius plot of $1/D$ hence obtained. 
When compared with amorphous water, we notice however that relative to the high temperature the temperature dependence of the dynamics at low temperature is rather weak, corresponding to an effective low-temperature energy barrier being too small relative to the high-temperature barrier.  
This issue can be attributed to a constant barrier offset $E_{0}$ adopted in \eq{w}.
\begin{figure}[tb]
\includegraphics[width=\columnwidth]{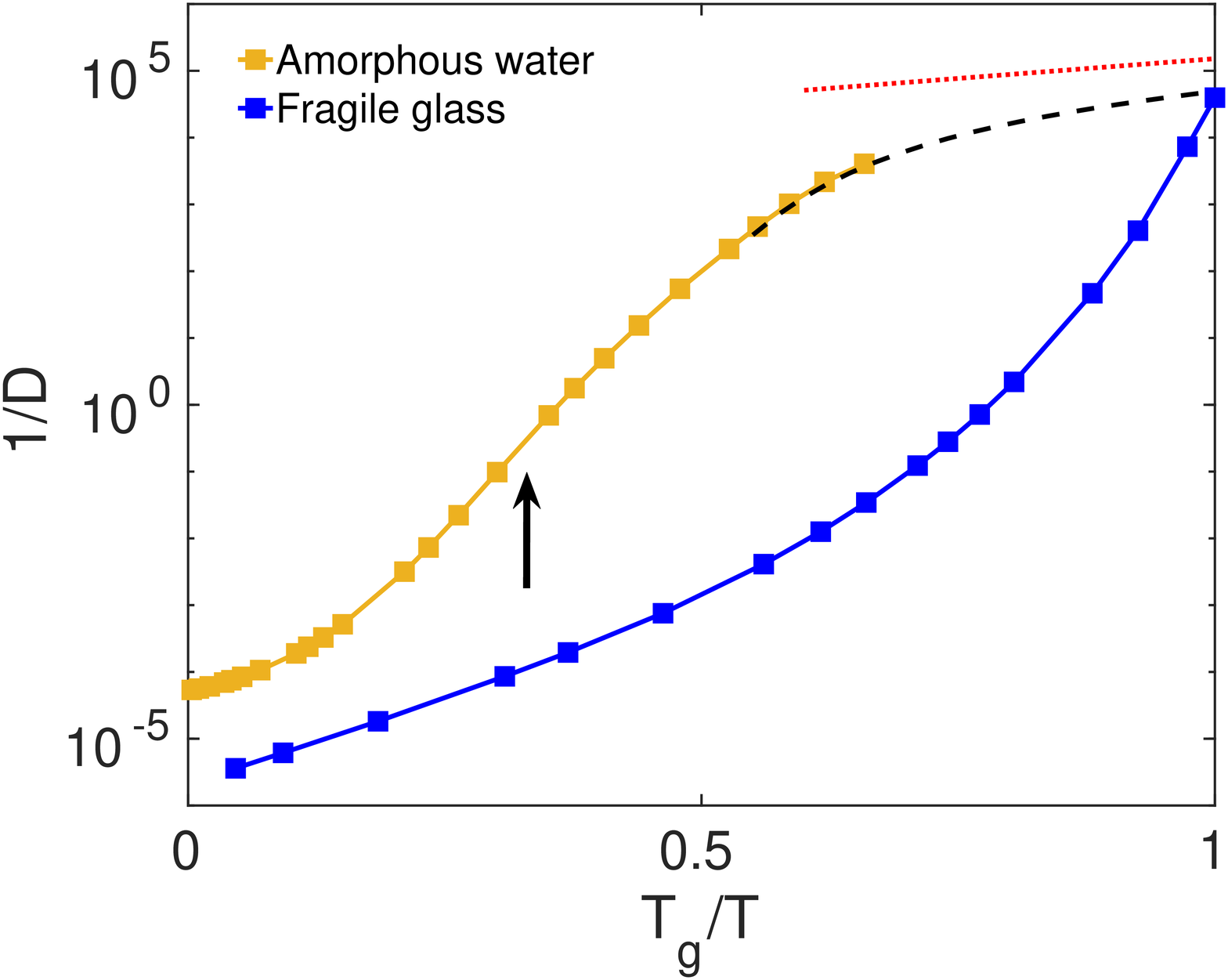}
\caption{Kinetic Angell plot for a model of amorphous water with a FTS transition and a fragile glass. For the FTS transition case, the dashed line represents a fit to \eq{ArrheniusFit2}. 
  The dotted line following \eq{Arrhenius2} is a guide to the eye.
  The arrow points to temperature $T=0.35$ at which $\beta$ is minimized.}
\label{angell_plot}
\end{figure}

In molecular systems, barriers of particle motions depend on the precise molecular structures and in particular on the bond types. A higher barrier is expected for breaking a hydrogen bond.  
We thus generalize the model to include an additional bond-dependent barrier offset term. Consider particle $k$ attempting to hop from  site \(i\) to site \(j\). Let \(\Omega_{ij}\) be the set of all other particles at any of the six nearest neighboring sites of \(i\) and \(j\). The rate in \eq{w} is generalized to
\begin{equation}
w  = w_{0}\exp{\squarebk{-\frac{E_{0}+\theta(\Delta E)~\Delta E+U(k,\Omega_{ij})}{k_{B}T}}}.
\label{metropolis_FTS}
\end{equation}
Here, we have included an additional hopping energy barrier term \(U(k,\Omega_{ij})\)  of magnitude $U_0$ defined as
\begin{equation}
U(k,\Omega_{ij}) = \frac{U_{0}}{C}\sum_{l \in \{\Omega_{ij}\}}^{} \delta (V_{kl},V_{0}),
\label{U}
\end{equation}
where $C=6$ represents the number of bonds affected by the hop,  including three to be broken and three to be formed. If all of these six bonds are all hydrogen bonds, \eq{U} gives $U(k,\Omega_{ij})=U_0$. This occurs at low temperature. In the other extreme, if all are non-hydrogen bonds, $U(k,\Omega_{ij})=0$, which may happen at high temperature.
Note that dynamics based on \eqs{metropolis_FTS}{U} satisfies detailed balance due to the symmetry \(U(k,\Omega_{ij})=U(k,\Omega_{ji})\) corresponding to the same barrier offset for forward and backward hops.  

To illustrate the physical relevance of $U_0$, \fig{vary_U0} also shows measured $1/D$ for various values of \(U_{0}\geq0 \). 
It can be observed that a larger $U_0$ leads to further slow down in the dynamics at low temperature, but it has relatively little effect on the dynamics at high temperature. Introducing $U_0$ thus allows the fine-tuning of the effective activation energy at low and high temperatures independently. 
	
To best reproduce dynamical features of amorphous water within reasonable computational requirements, we consider \(G_{1}\) = 0.38, \(U_{0}=0.4\), $E_0=0$ and \(w_{0}=10^4\). \Fig{angell_plot} shows a kinetic Angell plot of $1/D$ against $T_g/T$ thus obtained.  
Here, the glass transition temperature \(T_{g}\) is defined as the temperature at which \(1/D = 5\times10^4\), close to the slowest dynamic rates that simulations are still practical.
We obtain \(T_{g}=0.147\) after
extrapolating our simulation data to lower temperatures using 
\begin{equation}
\label{ArrheniusFit2}
D = D_0 \exp{\squarebk{ -\frac{E_0+U_0}{k_B T} + \roundbk{\frac{c_1}{T}-c_2}^{-4}}},
\end{equation}
which is an empirical form generalized from \eq{ArrheniusFit}. At low temperature, it reduces to
\begin{equation}
\label{Arrhenius2}
D \sim  \exp{\roundbk{ -\frac{E_0+U_0}{k_B T}}}. 
\end{equation}
which generalizes \eq{Arrhenius}.
For comparison, we also simulate a model of fragile glass 
by adopting the uniform-plus-delta interaction in \eq{gVfragile} with \(G_{0}=0.03\), \(E_{0}=2.05\), \(w_{0}=10^6\) and $U_0=0$.
We get $T_g=0.185$.
Results have also been shown in \fig{angell_plot}. We have chosen the model parameters so that the respective  features observed resemble qualitatively those of amorphous water and o-Terphenyl respectively as reported in \cite{shi2018}.

\begin{figure}[tb]
\includegraphics[width=\columnwidth]{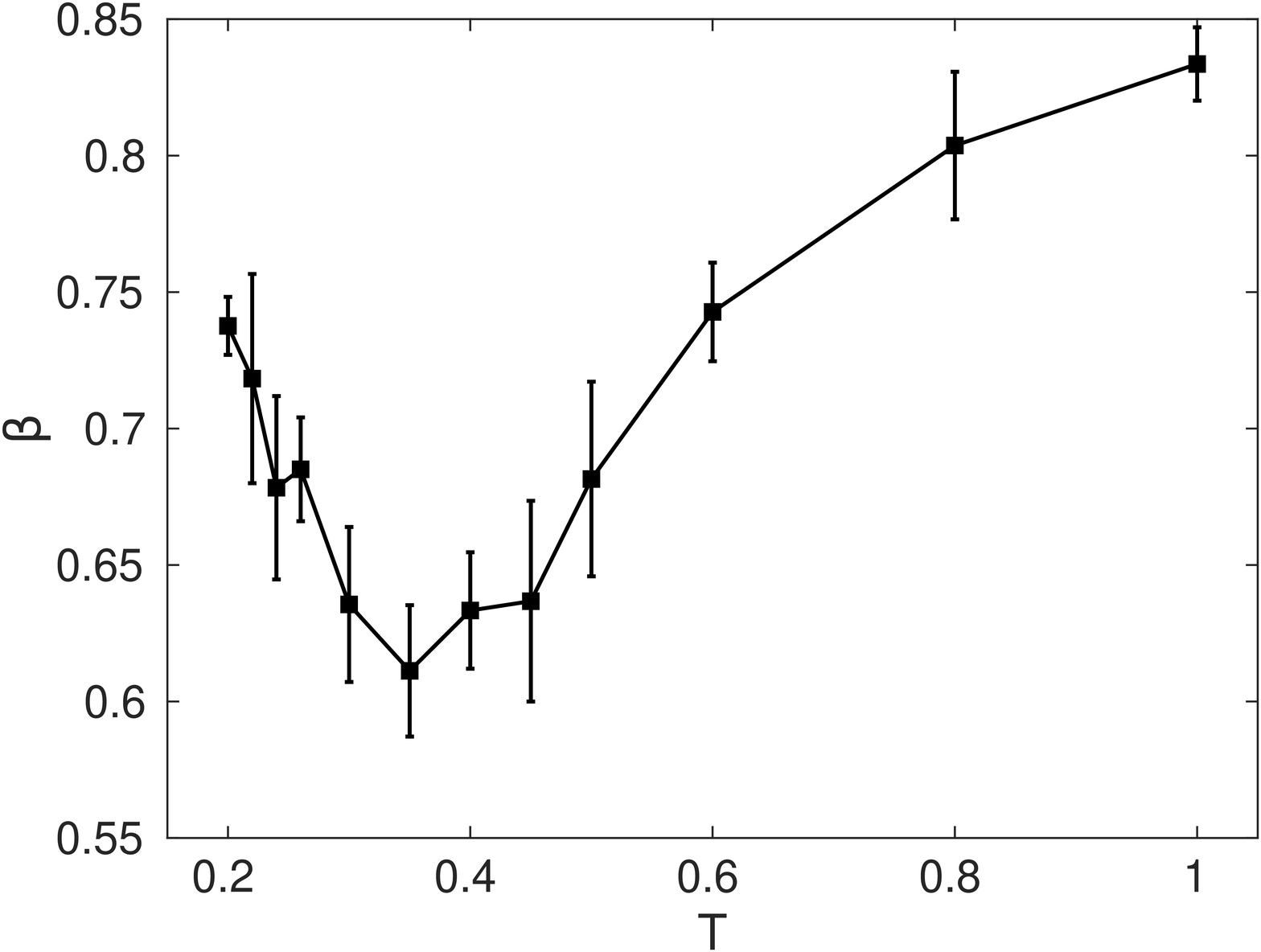}
\caption{Stretching exponent \(\beta\) plotted against \(T\) for the model of amorphous water. }
\label{beta}
\end{figure}

Besides diffusion coefficient, 
we have also measured the self-intermediate scattering function (see \App{app:methods}). It follows the Kohlrausch-Williams-Watts (KWW) form characterized by a stretching exponent $\beta$. \Fig{beta} plots the temperature dependence of $\beta$ obtained from the KWW fits. We observe a minimum of $\beta$ at \(T \simeq 0.35\).
This value is consistent with the temperature at which  the system starts to display the low temperature Arrhenius behavior as seen in \fig{angell_plot}. 
Similar observations have been reported in molecular dynamics simulations of TIP5P and ST2 models of water with the minimum $\beta$ associated with a maximized dynamic heterogeneity~\cite{shi2018,shi2020}.

\begin{figure}[tb]
\includegraphics[width=\columnwidth]{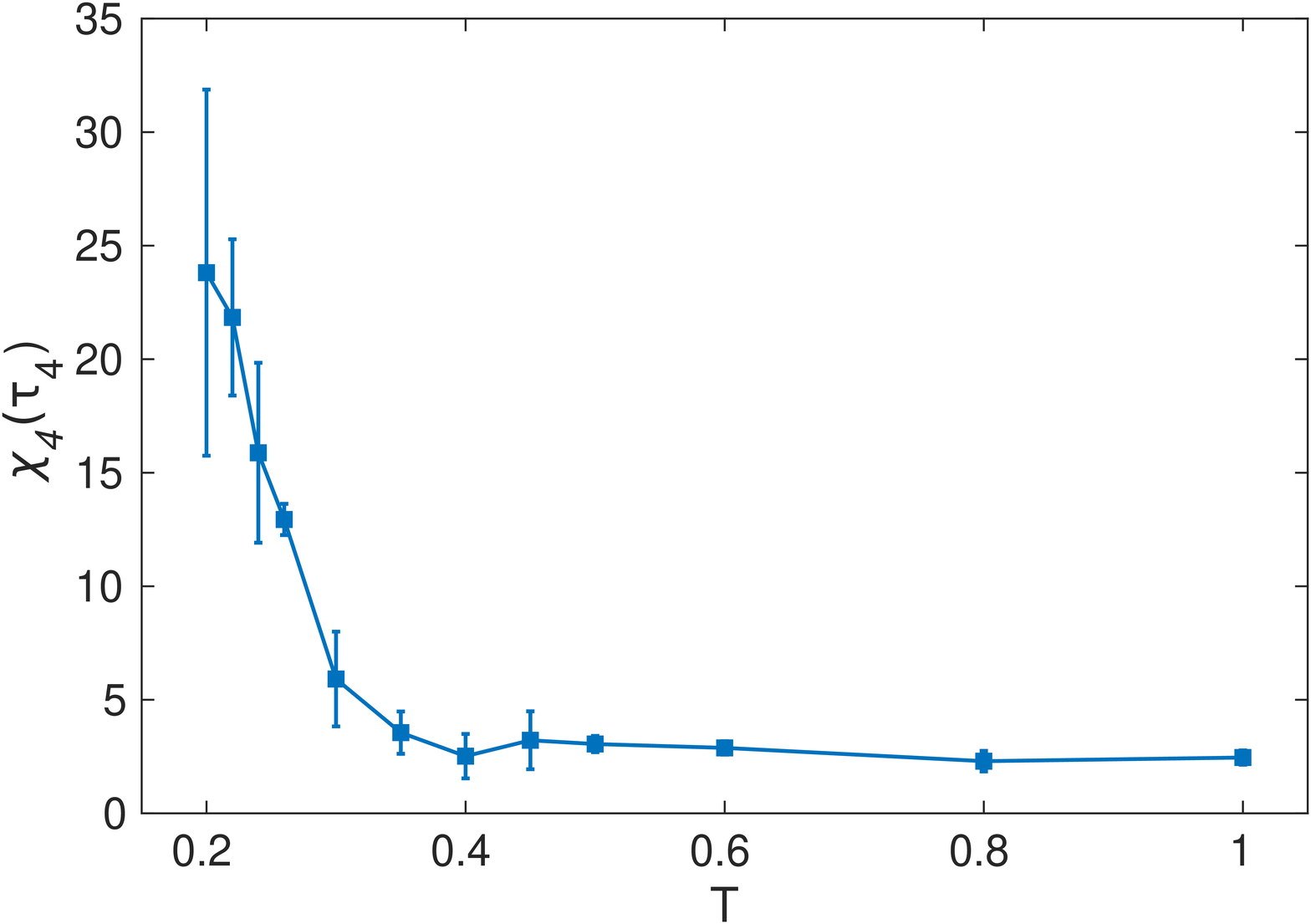}
\caption{Peak height $\chi_{4}(\tau_{4})$ of four-point correlation function plotted against  \(T\) for the model of amorphous water. 
}
\label{max_chi4}
\end{figure}

    In addition, we have measured the four-point correlation function \(\chi_{4}\) which shows a peak at time $\tau_{4}$  (see \App{app:methods}). \Fig{max_chi4} plots the peak height \(\chi_{4}(\tau_{4})\) as a function of temperature.  MD simulations in \cite{shi2018} shows that \(\chi_{4}(\tau_{4})\) exhibits a local maximum at the same temperature at which $\beta$ is minimized . However, we observe from \fig{max_chi4} no noticeable  maximization at $T\simeq 0.35$ at which $\beta$ is minimized.

\section{One-dimensional Model}

Our main simulations above have been performed in 2D. Many glassy features are believed to be independent of dimensions and previous 2D simulations of the DPLM are able to reproduce qualitatively many characteristic properties of glass \cite{zhang2017,lulli2020,lee2020,lee2021,gao2022,gopinath2022}. To show that our results are independent of dimensions, we perform additional study in 1D, which is computationally more efficient than in 2D. In a truely 1D lattice, particles cannot swap positions under the assumed void-induced dynamics and this can drastically alter the model characteristics. Our 1D model is therefore, more precisely, a quasi-1D model in which we shrink the width of the lattice in one direction and adopt a $L \times 2$ lattice with $L=600$. We apply periodic boundary conditions in both the long and short directions. Note that we keep taking $C=6$ in \eq{U} as a hop attempt may involve either five or six bonds. Other parts of the algorithm are unchanged. This system allows particles to swap positions via a sequence of void-induced hops. 

\begin{figure}[tb]
\includegraphics[width=\columnwidth]{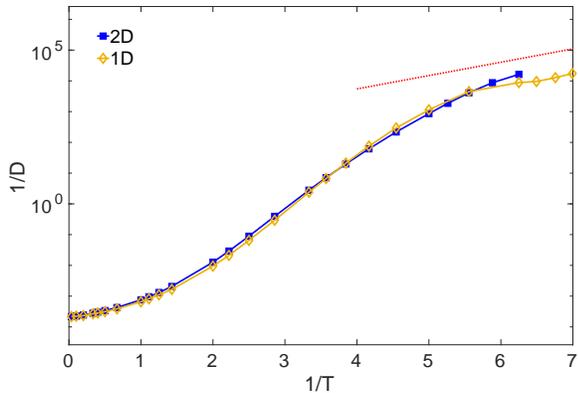}
\caption{
An Arrhenius plot of $1/D$ against $1/T$ for a 1D model of amorphous water with  $G_1=0.2$,  $U_0=1.5$, $w_0=6.7 \times 10^{5}$ and $\phi_v=0.03$. This is compared with results for a 2D model  with  $G_1=0.38$,  $U_0=0.4$,  $w_0=10^{4}$, and $\phi_v=0.01$. In both cases, we take $E_0=0$. The dotted line following \eq{Arrhenius2} is a guide to the eye.}

\label{1dsimulation}
\end{figure}

We consider $G_1=0.15$ and $w_0= 6.7\times 10^5$ so as to match the 2D results. We also increase the void density to $\phi_{v}=0.03$ from 0.01 so that average separation between voids is comparable to that in 2D. \Fig{1dsimulation} shows 1D results on $1/D$ against $1/T$. Compared with 2D results reproduced from \fig{angell_plot}, much similarity is observed. A wider low-temperature Arrhenius regime below the FTS transition can now be observed in 1D due to the better computational efficiency.

~\\
\section{Discussions}
\label{discussion}

As explained in \Cite{lee2020}, fragile glass can be produced by the DPLM using the uniform-plus-delta pair-interaction distribution $g(V)$ in \eq{gVfragile} as shown in \fig{angell_plot}. Closely related to a two-state picture of glass \cite{moynihan2000}, the two components in the distribution represents unexcited and excited states with average energies $V_0+k_BT$ and $V_1$ respectively. Glasses of the highest fragility is obtained when the weight $G_0$ of the unexcited state is small but non-vanishing. These unexcited interactions are thus rare and have a low entropy. As temperature decreases, interactions are increasingly restricted to these rare pairings, leading to a reduction of energetically favorable kinetic pathways and a dramatic slowdown of the dynamics. The unexcited component in $g(V)$ in \eq{gVfragile} has been approximated as a uniform distribution so that energy dispersion persists even when this component dominates at very low temperature. Super-Arrhenius behaviors thus persist to the lowest temperatures studied.

In this work, the fragile glass model is converted to a system with a FTS transition simply by replacing the lower uniform component of $g(V)$ by a delta function at $V_0$ in \eq{gVfts}. At high temperature, such details in the lower component have little impact because dynamics is dominated by whether individual interactions belong to the lower or the upper component. The fragile nature at high-temperature is thus preserved. At low temperature when the lower-energy component dominates, the non-dispersive energy at $V_0$ results at an Arrhenius behavior of strong glass as shown in \eq{Arrhenius2}. For amorphous water, the high strength of the hydrogen bonds may justify a relatively small dispersion in bond lengths and the bond energies despite the frustration.

Although there have been much progress in studying fragile glass and amorphous systems with FTS transitions, their relations are much less discussed. In particular, closely related two-state models \cite{moynihan2000,tanaka2000,tanaka2003}   have found great success in both their studies. Our lattice model approach allows studying both phenomena in a unified framework, their possible relations and differences in light of the two-state pictures are explicitly illustrated.

Lattice simulations are in general orders of magnitude faster than MD approaches, However, simulations at realistic time scales are still impossible and phenomena can often only be studied qualitatively at compressed time scales. Yet, we notice that the two characteristic turns in the Angell plot from our simulations are too stretched out.
More precisely, the FTS transition temperature and the most super-Arrhenius point  from \fig{angell_plot} occurs roughly at $T_g/T \simeq$ 0.42 and 0.15 respectively, corresponding to a ratio of about 2.8. This is much larger than a ratio of roughly 1.5 from MD simulations \cite{shi2018}. We believe that the discrepancy may be due to a lack of correlation between neighboring bonds in our model, which is a feature of the DPLM \cite{zhang2017} and interestingly is also assumed implicitly in two-state models \cite{moynihan2000,tanaka2000,tanaka2003}. 

    To conclude, using a lattice model of glass, we have reproduced a FTS transition separating a high-temperature fragile phase from a low-temperature strong phase. 
The model differs from that of fragile glass mainly by a different low-energy component of the pair-interaction energy distribution. While highly dispersed low-energy states lead to a fragile glass for the whole temperature range, a narrow low-energy distribution induces a FTS transition to a low-temperature strong regime. Our work provides a unified framework for studying fragile glass and systems exhibiting FTS transitions.

\acknowledgements

We thank helpful discussions with Rui Shi. This work was supported by General Research Fund of Hong Kong (Grant 15303220) and National Natural Science Foundation of China (Grant 11974297).
~\\~\\~\\~\\~
~\\~\\~\\~\\~
~\\~\\~\\~\\~
~\\~\\~\\~\\~
~\\~\\~\\~\\~
~\\~\\~\\~\\~
~\\~\\~\\~\\~
\appendix

\setcounter{figure}{0}
\renewcommand{\figurename}{Fig.}
\renewcommand{\thefigure}{S\arabic{figure}}
\section{Details of dynamical characterization}
\label{app:methods}
\begin{figure}[t]
	\includegraphics[width=\columnwidth]{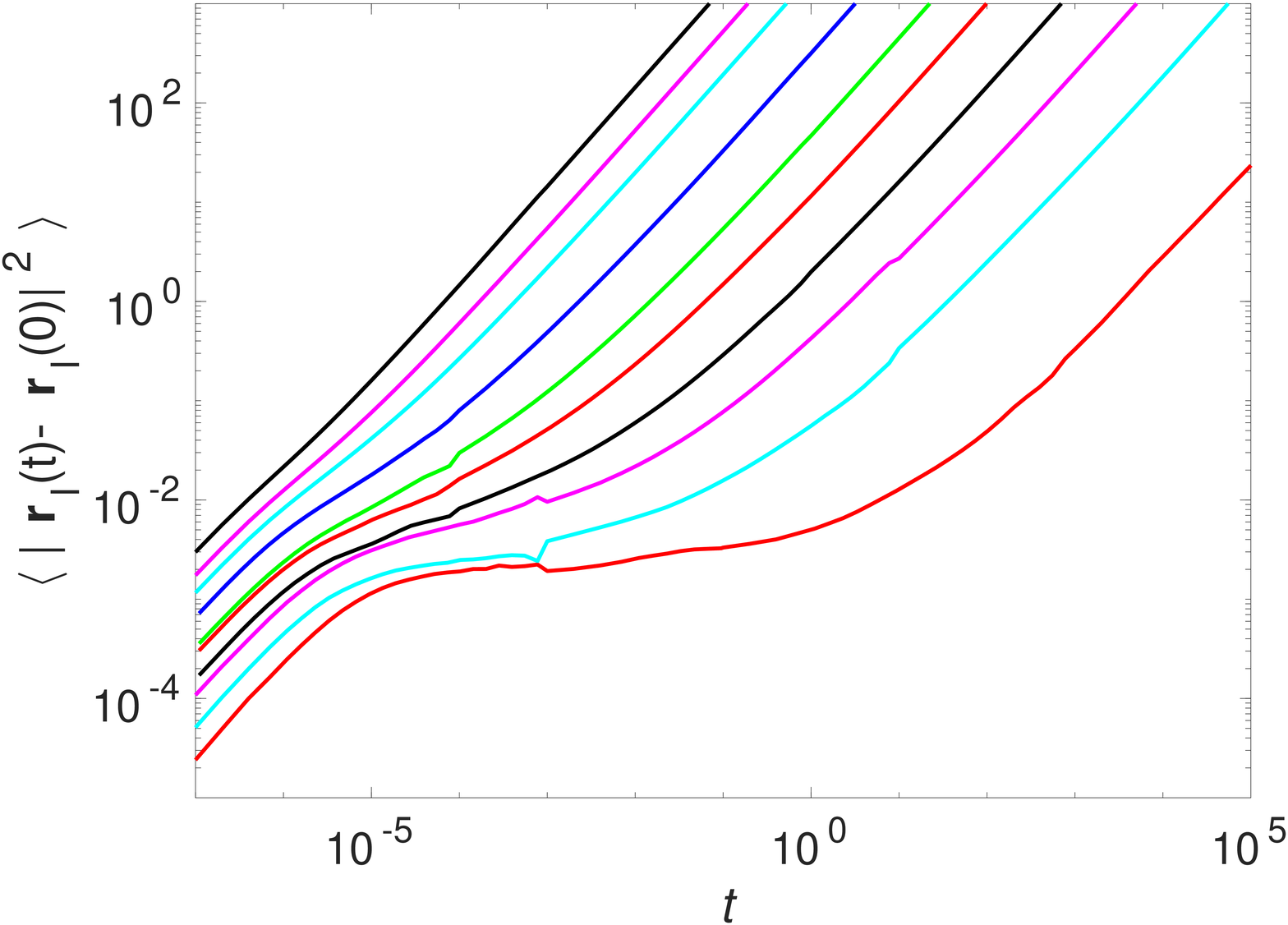}
	\caption{MSD against time $t$ for \(T\) = 3, 1, 0.7, 0.5, 0.4, 0.35, 0.3, 0.26, 0.22, 0.16   (from left to right).}
	\label{MSD}
\end{figure}
\begin{figure}[tb]
	\includegraphics[width=\columnwidth]{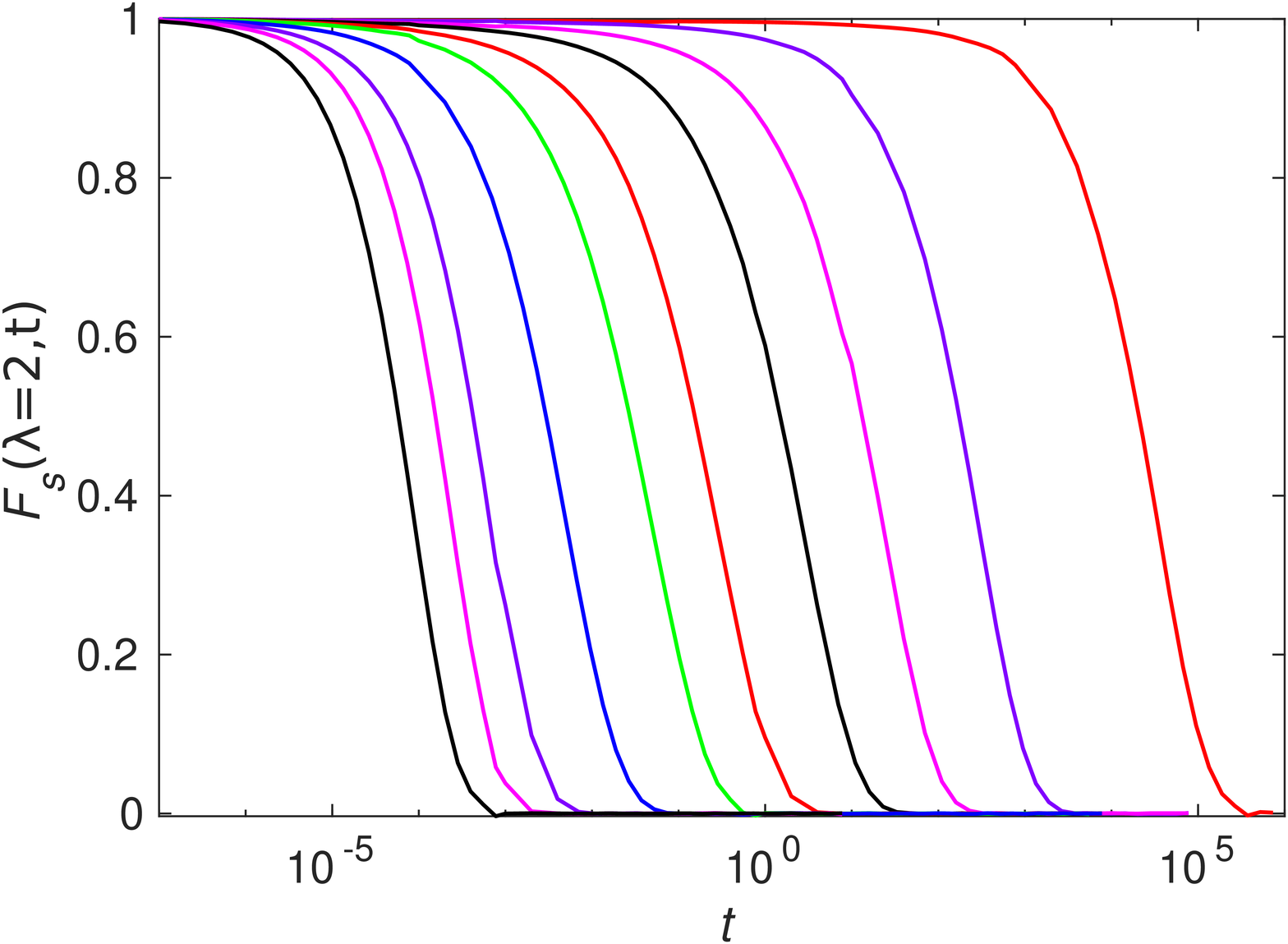}
	\caption{Self-intermediate scattering function \(F_{s}(\textbf{q},t)\) with $\av{\textbf{q}} = 2\pi/\lambda$ and $\lambda$ = 2 against time $t$ for \(T\) = 3, 1, 0.7, 0.5, 0.4, 0.35, 0.3, 0.26, 0.22, 0.16 (from left to right). }
	\label{Fs}
\end{figure}
We now provide full details in the characterization of the dynamics of the DPLM with a bi-delta pair-interaction distribution $g(V)$ in \eq{gVfts} which, shows a FTS transition.
We will take the example of our 2D model of water introduced in Sec. \ref{water}. Results for other parameters are qualitatively similar.
As already explained, we adopts a probabilistic weight $G_1=0.38$ of the low-energy component. Particle hops follow \eq{metropolis_FTS} with $E_0=0$, $w_0=10^6$ and $U_0=0.4$. A void density of $\phi_v=0.01$ is used. 

We have calculated the particle mean square displacement (MSD) from \(\big\langle|\textbf{r}_{l}(t) – \textbf{r}_{l}(0)|^2\big\rangle\) with $\textbf{r}_{l}(t)$ denoting the position of particle $l$ at time $t$. \Fig{MSD} plots the MSD against time \(t\). We observe the development of a plateau as temperature decreases, The particle diffusion coefficient $D$ is then calculated from \(D = (1/2d)(\text{MSD}/t)\) at long time $t$ and results are plotted in \fig{angell_plot}. Here, \(d=2\) represents the dimension of the system. To ensure a sufficiently long time $t$ in the calculation, only MSD values beyond 1  satisfying $MSD \sim t^\gamma$ with $\gamma \ge 0.96$ are considered.

\begin{figure}[tb]
\includegraphics[width=\columnwidth]{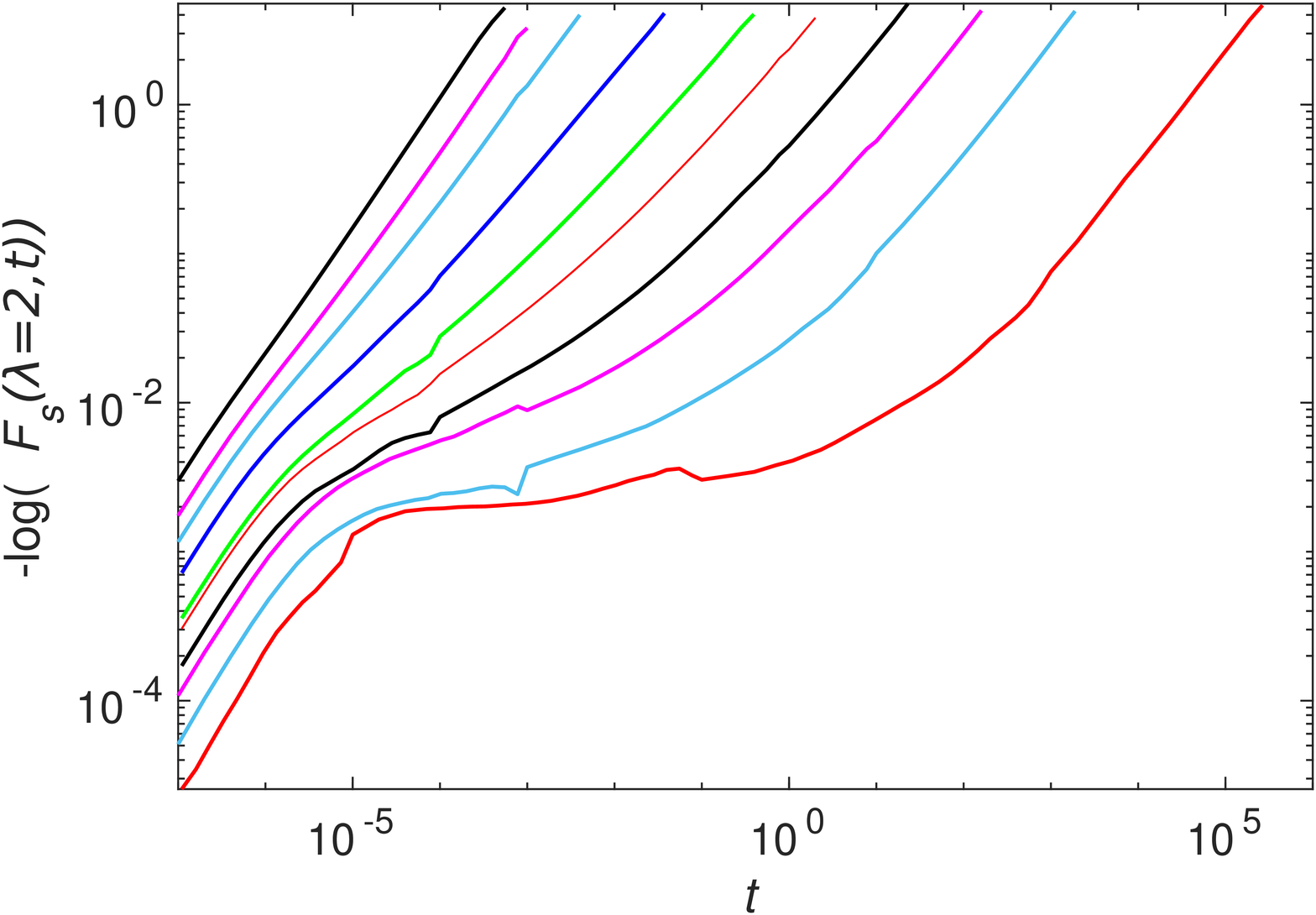}
\caption{Minus-log of self-intermediate scattering function against time $t$ resulting at a log-log-versus-log plot  using data from \fig{Fs}.}

\label{logFs}
\end{figure}

\begin{figure}[tb]
\includegraphics[width=\columnwidth]{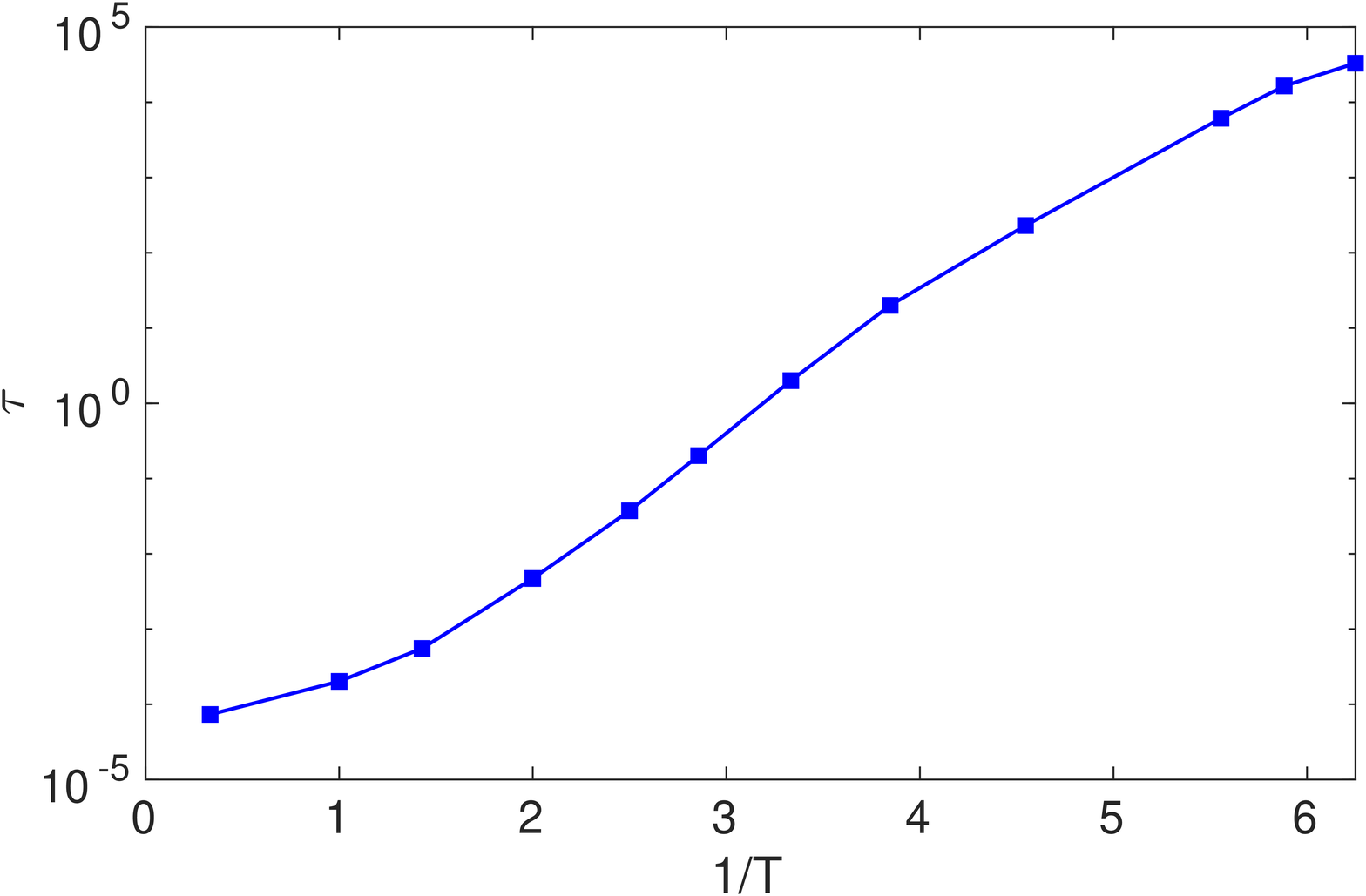}
\caption{Structural relaxation time $\tau$ against $1/T$. }
\label{tau}
\end{figure}

\begin{figure}[h]
\includegraphics[width=\columnwidth]{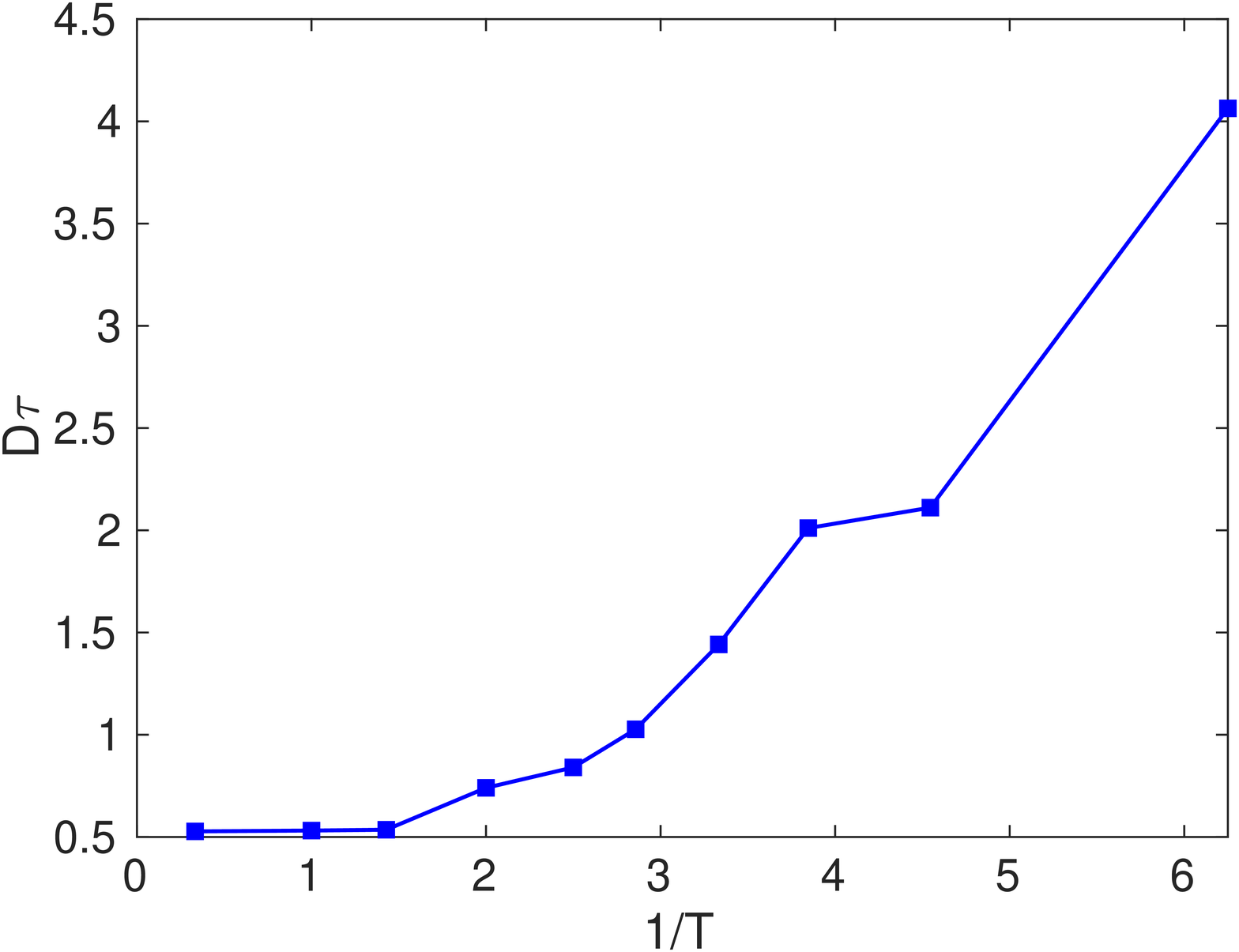}
\caption{Plot of $D\tau$ against $1/T$. The non-constant values show a violation of Stokes-Einstein relation.}
\label{stokes_einstein}
\end{figure}

Next, we examine 
the self-intermediate scattering function \(F_{s}\)
defined as
\begin{equation}
F_{s}(\textbf{q},t)=\bigg \langle e^{i \textbf{q} \cdot (\textbf{r}_{l}(t) – \textbf{r}_{l}(0))}\bigg \rangle,
\end{equation}
where $\av{\textbf{q}} = 2\pi/\lambda$ with $\lambda$ = 2. \fig{Fs} plots \(F_{s}\) against \(t\). It shows a two-step relaxation with a tiny first step. The main relaxation in $F_s$ follows the Kohlrausch-Williams-Watts (KWW) stretched exponential function \(A\exp(-(t/\tau)^\beta)\), where $\beta$ is the stretching exponent, $\tau$ is the relaxation time and A is a constant close to 1. We plot $-\log(F_{s})$ against \(t\) in log-log scales in \fig{logFs} from which we obtain $\beta$ for different \(T\) by computing the slope of the linear region at large \(t\) satisfying \(10^{-3}\leq F_{s}(\textbf{q},t) \leq 0.9\). Results on $\beta$ are shown in \fig{beta}.
The relaxation time $\tau$ can be obtained as the time at which $F_{s}=1/e$.
In \fig{tau}, we plot \(\tau\) against \(1/T\) . \Fig{stokes_einstein} then further show a violation in Stokes-Einstein relation as observed from an increase of \(D\tau\) when \(T\)  decreases.
\begin{figure}[tb]
	\includegraphics[width=\columnwidth]{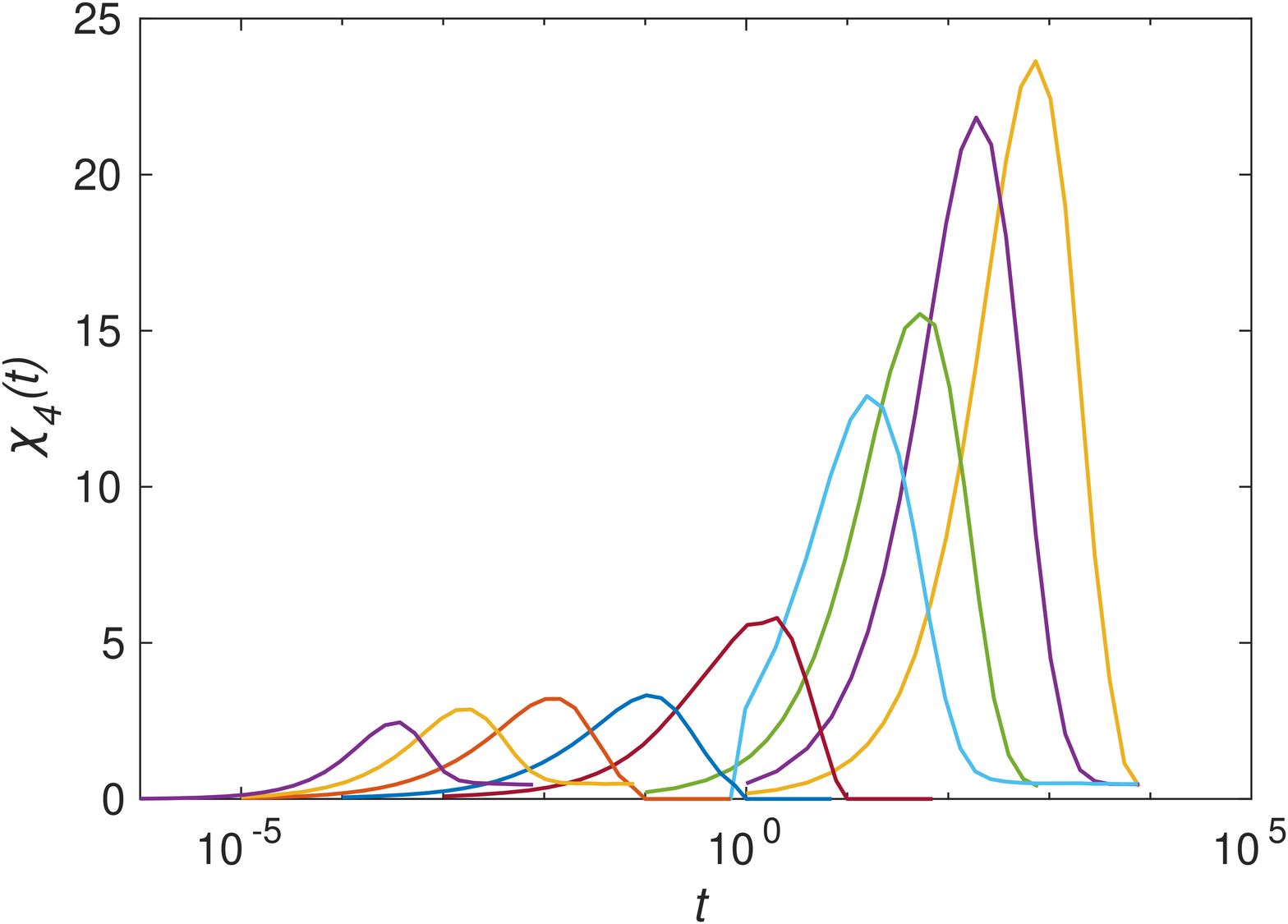}
	\caption{Four-point correlation function $\chi_{4}(t)$ against time $t$ for \(T\) = 1, 0.6, 0.45, 0.35, 0.3, 0.26, 0.24, 0.22 and 0.20 (from left to right).}
	\label{chi4}
\end{figure}
We have also studied the four-point correlation function defined by	\(\chi_{4}(t) = \lim_{\textbf{\~q}\to\ 0} S_{4}(\textbf{\~{q}},t)\), where
\begin{equation}
 S_{4}(\textbf{\~{q}},t)=\frac{1}{N} \bigg\langle \bigg|\sum_{l}{}e^{i\textbf{\~{q}} \cdot \textbf{r}_{l}(0)}
\squarebk{e^{i\textbf{{q}} \cdot(\textbf{r}_{l}(t) – \textbf{r}_{l}(0))}-F_{s}(\textbf{q},t)}
      \bigg|^2 \bigg\rangle.
\end{equation}
\Fig{chi4} plots \(\chi_{4}(t)\) against time \(t\) for various \(T\). We observe that \(\chi_{4}(t)\) maximizes at a dynamical timescale $\tau_{4}$. The peak height \(\chi_{4}(\tau_4)\) is plotted against $T$ in \fig{max_chi4}.

~\\~\\~\\~\\~
~\\~\\~\\~\\~
~\\~\\~\\~\\~
~\\~\\~\\~\\~
~\\~\\~\\~\\~

%


\begin{thebibliography}{31}%
	\makeatletter
	\providecommand \@ifxundefined [1]{%
		\@ifx{#1\undefined}
	}%
	\providecommand \@ifnum [1]{%
		\ifnum #1\expandafter \@firstoftwo
		\else \expandafter \@secondoftwo
		\fi
	}%
	\providecommand \@ifx [1]{%
		\ifx #1\expandafter \@firstoftwo
		\else \expandafter \@secondoftwo
		\fi
	}%
	\providecommand \natexlab [1]{#1}%
	\providecommand \enquote  [1]{``#1''}%
	\providecommand \bibnamefont  [1]{#1}%
	\providecommand \bibfnamefont [1]{#1}%
	\providecommand \citenamefont [1]{#1}%
	\providecommand \href@noop [0]{\@secondoftwo}%
	\providecommand \href [0]{\begingroup \@sanitize@url \@href}%
	\providecommand \@href[1]{\@@startlink{#1}\@@href}%
	\providecommand \@@href[1]{\endgroup#1\@@endlink}%
	\providecommand \@sanitize@url [0]{\catcode `\\12\catcode `\$12\catcode
		`\&12\catcode `\#12\catcode `\^12\catcode `\_12\catcode `\%12\relax}%
	\providecommand \@@startlink[1]{}%
	\providecommand \@@endlink[0]{}%
	\providecommand \url  [0]{\begingroup\@sanitize@url \@url }%
	\providecommand \@url [1]{\endgroup\@href {#1}{\urlprefix }}%
	\providecommand \urlprefix  [0]{URL }%
	\providecommand \Eprint [0]{\href }%
	\providecommand \doibase [0]{https://doi.org/}%
	\providecommand \selectlanguage [0]{\@gobble}%
	\providecommand \bibinfo  [0]{\@secondoftwo}%
	\providecommand \bibfield  [0]{\@secondoftwo}%
	\providecommand \translation [1]{[#1]}%
	\providecommand \BibitemOpen [0]{}%
	\providecommand \bibitemStop [0]{}%
	\providecommand \bibitemNoStop [0]{.\EOS\space}%
	\providecommand \EOS [0]{\spacefactor3000\relax}%
	\providecommand \BibitemShut  [1]{\csname bibitem#1\endcsname}%
	\let\auto@bib@innerbib\@empty
	\bibitem [{\citenamefont {Stillinger}\ and\ \citenamefont
		{Debenedetti}(2013)}]{stillinger2013review}%
	\BibitemOpen
	\bibfield  {author} {\bibinfo {author} {\bibfnamefont {F.~H.}\ \bibnamefont
			{Stillinger}}\ and\ \bibinfo {author} {\bibfnamefont {P.~G.}\ \bibnamefont
			{Debenedetti}},\ }\bibfield  {title} {\bibinfo {title} {Glass transition
			thermodynamics and kinetics},\ }\href
	{https://doi.org/10.1146/annurev-conmatphys-030212-184329} {\bibfield
		{journal} {\bibinfo  {journal} {Annu. Rev. Condens. Matter Phys.}\ }\textbf
		{\bibinfo {volume} {4}},\ \bibinfo {pages} {263} (\bibinfo {year}
		{2013})}\BibitemShut {NoStop}%
	\bibitem [{\citenamefont {Biroli}\ and\ \citenamefont
		{Garrahan}(2013)}]{biroli2013review}%
	\BibitemOpen
	\bibfield  {author} {\bibinfo {author} {\bibfnamefont {G.}~\bibnamefont
			{Biroli}}\ and\ \bibinfo {author} {\bibfnamefont {J.~P.}\ \bibnamefont
			{Garrahan}},\ }\bibfield  {title} {\bibinfo {title} {Perspective: The glass
			transition},\ }\href@noop {} {\bibfield  {journal} {\bibinfo  {journal} {J.
				Chem. Phys.}\ }\textbf {\bibinfo {volume} {138}},\ \bibinfo {pages} {12A301}
		(\bibinfo {year} {2013})}\BibitemShut {NoStop}%
	\bibitem [{\citenamefont {Arceri}\ \emph {et~al.}(2020)\citenamefont {Arceri},
		\citenamefont {Landes}, \citenamefont {Berthier},\ and\ \citenamefont
		{Biroli}}]{arceri2020}%
	\BibitemOpen
	\bibfield  {author} {\bibinfo {author} {\bibfnamefont {F.}~\bibnamefont
			{Arceri}}, \bibinfo {author} {\bibfnamefont {F.~P.}\ \bibnamefont {Landes}},
		\bibinfo {author} {\bibfnamefont {L.}~\bibnamefont {Berthier}},\ and\
		\bibinfo {author} {\bibfnamefont {G.}~\bibnamefont {Biroli}},\ }\bibfield
	{title} {\bibinfo {title} {Glasses and aging: A statistical mechanics
			perspective},\ }\href@noop {} {\bibfield  {journal} {\bibinfo  {journal}
			{arXiv:2006.09725}\ } (\bibinfo {year} {2020})}\BibitemShut {NoStop}%
	\bibitem [{\citenamefont {Ito}\ \emph {et~al.}(1999)\citenamefont {Ito},
		\citenamefont {Moynihan},\ and\ \citenamefont {Angell}}]{ito1999}%
	\BibitemOpen
	\bibfield  {author} {\bibinfo {author} {\bibfnamefont {K.}~\bibnamefont
			{Ito}}, \bibinfo {author} {\bibfnamefont {C.~T.}\ \bibnamefont {Moynihan}},\
		and\ \bibinfo {author} {\bibfnamefont {C.~A.}\ \bibnamefont {Angell}},\
	}\bibfield  {title} {\bibinfo {title} {Thermodynamic determination of
			fragility in liquids and a fragile-to-strong liquid transition in water},\
	}\href@noop {} {\bibfield  {journal} {\bibinfo  {journal} {Nature}\ }\textbf
		{\bibinfo {volume} {398}},\ \bibinfo {pages} {492} (\bibinfo {year}
		{1999})}\BibitemShut {NoStop}%
	\bibitem [{\citenamefont {Angell}(1993)}]{angell1993}%
	\BibitemOpen
	\bibfield  {author} {\bibinfo {author} {\bibfnamefont {C.~A.}\ \bibnamefont
			{Angell}},\ }\bibfield  {title} {\bibinfo {title} {Water ii is a "strong"
			liquid},\ }\href@noop {} {\bibfield  {journal} {\bibinfo  {journal} {J. Phys.
				Chem.}\ }\textbf {\bibinfo {volume} {97}},\ \bibinfo {pages} {6339} (\bibinfo
		{year} {1993})}\BibitemShut {NoStop}%
	\bibitem [{\citenamefont {Saksaengwijit}\ \emph {et~al.}(2004)\citenamefont
		{Saksaengwijit}, \citenamefont {Reinisch},\ and\ \citenamefont
		{Heuer}}]{saksaengwijit2004}%
	\BibitemOpen
	\bibfield  {author} {\bibinfo {author} {\bibfnamefont {A.}~\bibnamefont
			{Saksaengwijit}}, \bibinfo {author} {\bibfnamefont {J.}~\bibnamefont
			{Reinisch}},\ and\ \bibinfo {author} {\bibfnamefont {A.}~\bibnamefont
			{Heuer}},\ }\bibfield  {title} {\bibinfo {title} {Origin of the
			fragile-to-strong crossover in liquid silica as expressed by its
			potential-energy landscape},\ }\href@noop {} {\bibfield  {journal} {\bibinfo
			{journal} {Phys. Rev. Lett.}\ }\textbf {\bibinfo {volume} {93}},\ \bibinfo
		{pages} {235701} (\bibinfo {year} {2004})}\BibitemShut {NoStop}%
	\bibitem [{\citenamefont {Hemmati}\ \emph {et~al.}(2001)\citenamefont
		{Hemmati}, \citenamefont {Moynihan},\ and\ \citenamefont
		{Angell}}]{hemmati2001}%
	\BibitemOpen
	\bibfield  {author} {\bibinfo {author} {\bibfnamefont {M.}~\bibnamefont
			{Hemmati}}, \bibinfo {author} {\bibfnamefont {C.~T.}\ \bibnamefont
			{Moynihan}},\ and\ \bibinfo {author} {\bibfnamefont {C.~A.}\ \bibnamefont
			{Angell}},\ }\bibfield  {title} {\bibinfo {title} {Interpretation of the
			molten bef 2 viscosity anomaly in terms of a high temperature density
			maximum, and other waterlike features},\ }\href@noop {} {\bibfield  {journal}
		{\bibinfo  {journal} {J. Chem. Phys.}\ }\textbf {\bibinfo {volume} {115}},\
		\bibinfo {pages} {6663} (\bibinfo {year} {2001})}\BibitemShut {NoStop}%
	\bibitem [{\citenamefont {Evenson}\ \emph {et~al.}(2012)\citenamefont
		{Evenson}, \citenamefont {Schmitt}, \citenamefont {Nicola}, \citenamefont
		{Gallino},\ and\ \citenamefont {Busch}}]{evenson2012}%
	\BibitemOpen
	\bibfield  {author} {\bibinfo {author} {\bibfnamefont {Z.}~\bibnamefont
			{Evenson}}, \bibinfo {author} {\bibfnamefont {T.}~\bibnamefont {Schmitt}},
		\bibinfo {author} {\bibfnamefont {M.}~\bibnamefont {Nicola}}, \bibinfo
		{author} {\bibfnamefont {I.}~\bibnamefont {Gallino}},\ and\ \bibinfo {author}
		{\bibfnamefont {R.}~\bibnamefont {Busch}},\ }\bibfield  {title} {\bibinfo
		{title} {High temperature melt viscosity and fragile to strong transition in
			zr--cu--ni--al--nb (ti) and cu47ti34zr11ni8 bulk metallic glasses},\
	}\href@noop {} {\bibfield  {journal} {\bibinfo  {journal} {Acta materialia}\
		}\textbf {\bibinfo {volume} {60}},\ \bibinfo {pages} {4712} (\bibinfo {year}
		{2012})}\BibitemShut {NoStop}%
	\bibitem [{\citenamefont {Gallo}\ \emph {et~al.}(2000)\citenamefont {Gallo},
		\citenamefont {Rovere},\ and\ \citenamefont {Spohr}}]{gallo2000}%
	\BibitemOpen
	\bibfield  {author} {\bibinfo {author} {\bibfnamefont {P.}~\bibnamefont
			{Gallo}}, \bibinfo {author} {\bibfnamefont {M.}~\bibnamefont {Rovere}},\ and\
		\bibinfo {author} {\bibfnamefont {E.}~\bibnamefont {Spohr}},\ }\bibfield
	{title} {\bibinfo {title} {Supercooled confined water and the mode coupling
			crossover temperature},\ }\href@noop {} {\bibfield  {journal} {\bibinfo
			{journal} {Phys. Rev. Lett.}\ }\textbf {\bibinfo {volume} {85}},\ \bibinfo
		{pages} {4317} (\bibinfo {year} {2000})}\BibitemShut {NoStop}%
	\bibitem [{\citenamefont {Xu}\ \emph {et~al.}(2005)\citenamefont {Xu},
		\citenamefont {Kumar}, \citenamefont {Buldyrev}, \citenamefont {Chen},
		\citenamefont {Poole}, \citenamefont {Sciortino},\ and\ \citenamefont
		{Stanley}}]{xu2005}%
	\BibitemOpen
	\bibfield  {author} {\bibinfo {author} {\bibfnamefont {L.}~\bibnamefont
			{Xu}}, \bibinfo {author} {\bibfnamefont {P.}~\bibnamefont {Kumar}}, \bibinfo
		{author} {\bibfnamefont {S.~V.}\ \bibnamefont {Buldyrev}}, \bibinfo {author}
		{\bibfnamefont {S.-H.}\ \bibnamefont {Chen}}, \bibinfo {author}
		{\bibfnamefont {P.~H.}\ \bibnamefont {Poole}}, \bibinfo {author}
		{\bibfnamefont {F.}~\bibnamefont {Sciortino}},\ and\ \bibinfo {author}
		{\bibfnamefont {H.~E.}\ \bibnamefont {Stanley}},\ }\bibfield  {title}
	{\bibinfo {title} {Relation between the widom line and the dynamic crossover
			in systems with a liquid--liquid phase transition},\ }\href@noop {}
	{\bibfield  {journal} {\bibinfo  {journal} {Proc. Natl. Acad. Sci.}\ }\textbf
		{\bibinfo {volume} {102}},\ \bibinfo {pages} {16558} (\bibinfo {year}
		{2005})}\BibitemShut {NoStop}%
	\bibitem [{\citenamefont {Tanaka}(2000)}]{tanaka2000}%
	\BibitemOpen
	\bibfield  {author} {\bibinfo {author} {\bibfnamefont {H.}~\bibnamefont
			{Tanaka}},\ }\bibfield  {title} {\bibinfo {title} {Simple physical model of
			liquid water},\ }\href@noop {} {\bibfield  {journal} {\bibinfo  {journal} {J.
				Chem. Phys.}\ }\textbf {\bibinfo {volume} {112}},\ \bibinfo {pages} {799}
		(\bibinfo {year} {2000})}\BibitemShut {NoStop}%
	\bibitem [{\citenamefont {Tanaka}(2003)}]{tanaka2003}%
	\BibitemOpen
	\bibfield  {author} {\bibinfo {author} {\bibfnamefont {H.}~\bibnamefont
			{Tanaka}},\ }\bibfield  {title} {\bibinfo {title} {A new scenario of the
			apparent fragile-to-strong transition in tetrahedral liquids: Water as an
			example},\ }\href@noop {} {\bibfield  {journal} {\bibinfo  {journal} {Journal
				of Physics: Condensed Matter}\ }\textbf {\bibinfo {volume} {15}},\ \bibinfo
		{pages} {L703} (\bibinfo {year} {2003})}\BibitemShut {NoStop}%
	\bibitem [{\citenamefont {Myneni}\ \emph {et~al.}(2002)\citenamefont {Myneni},
		\citenamefont {Luo}, \citenamefont {N{\"a}slund}, \citenamefont {Cavalleri},
		\citenamefont {Ojam{\"a}e}, \citenamefont {Ogasawara}, \citenamefont
		{Pelmenschikov}, \citenamefont {Wernet}, \citenamefont {V{\"a}terlein},
		\citenamefont {Heske} \emph {et~al.}}]{myneni2002}%
	\BibitemOpen
	\bibfield  {author} {\bibinfo {author} {\bibfnamefont {S.}~\bibnamefont
			{Myneni}}, \bibinfo {author} {\bibfnamefont {Y.}~\bibnamefont {Luo}},
		\bibinfo {author} {\bibfnamefont {L.~{\AA}.}\ \bibnamefont {N{\"a}slund}},
		\bibinfo {author} {\bibfnamefont {M.}~\bibnamefont {Cavalleri}}, \bibinfo
		{author} {\bibfnamefont {L.}~\bibnamefont {Ojam{\"a}e}}, \bibinfo {author}
		{\bibfnamefont {H.}~\bibnamefont {Ogasawara}}, \bibinfo {author}
		{\bibfnamefont {A.}~\bibnamefont {Pelmenschikov}}, \bibinfo {author}
		{\bibfnamefont {P.}~\bibnamefont {Wernet}}, \bibinfo {author} {\bibfnamefont
			{P.}~\bibnamefont {V{\"a}terlein}}, \bibinfo {author} {\bibfnamefont
			{C.}~\bibnamefont {Heske}}, \emph {et~al.},\ }\bibfield  {title} {\bibinfo
		{title} {Spectroscopic probing of local hydrogen-bonding structures in liquid
			water},\ }\href@noop {} {\bibfield  {journal} {\bibinfo  {journal} {Journal
				of Physics: Condensed Matter}\ }\textbf {\bibinfo {volume} {14}},\ \bibinfo
		{pages} {L213} (\bibinfo {year} {2002})}\BibitemShut {NoStop}%
	\bibitem [{\citenamefont {Nilsson}\ \emph {et~al.}(2010)\citenamefont
		{Nilsson}, \citenamefont {Nordlund}, \citenamefont {Waluyo}, \citenamefont
		{Huang}, \citenamefont {Ogasawara}, \citenamefont {Kaya}, \citenamefont
		{Bergmann}, \citenamefont {N{\"a}slund}, \citenamefont {{\"O}str{\"o}m},
		\citenamefont {Wernet} \emph {et~al.}}]{nilsson2010}%
	\BibitemOpen
	\bibfield  {author} {\bibinfo {author} {\bibfnamefont {A.}~\bibnamefont
			{Nilsson}}, \bibinfo {author} {\bibfnamefont {D.}~\bibnamefont {Nordlund}},
		\bibinfo {author} {\bibfnamefont {I.}~\bibnamefont {Waluyo}}, \bibinfo
		{author} {\bibfnamefont {N.}~\bibnamefont {Huang}}, \bibinfo {author}
		{\bibfnamefont {H.}~\bibnamefont {Ogasawara}}, \bibinfo {author}
		{\bibfnamefont {S.}~\bibnamefont {Kaya}}, \bibinfo {author} {\bibfnamefont
			{U.}~\bibnamefont {Bergmann}}, \bibinfo {author} {\bibfnamefont {L.-{\AA}.}\
			\bibnamefont {N{\"a}slund}}, \bibinfo {author} {\bibfnamefont
			{H.}~\bibnamefont {{\"O}str{\"o}m}}, \bibinfo {author} {\bibfnamefont
			{P.}~\bibnamefont {Wernet}}, \emph {et~al.},\ }\bibfield  {title} {\bibinfo
		{title} {X-ray absorption spectroscopy and x-ray raman scattering of water
			and ice; an experimental view},\ }\href@noop {} {\bibfield  {journal}
		{\bibinfo  {journal} {Journal of Electron Spectroscopy and Related
				Phenomena}\ }\textbf {\bibinfo {volume} {177}},\ \bibinfo {pages} {99}
		(\bibinfo {year} {2010})}\BibitemShut {NoStop}%
	\bibitem [{\citenamefont {Zhang}\ and\ \citenamefont {Lam}(2017)}]{zhang2017}%
	\BibitemOpen
	\bibfield  {author} {\bibinfo {author} {\bibfnamefont {L.-H.}\ \bibnamefont
			{Zhang}}\ and\ \bibinfo {author} {\bibfnamefont {C.-H.}\ \bibnamefont
			{Lam}},\ }\bibfield  {title} {\bibinfo {title} {Emergent facilitation
			behavior in a distinguishable-particle lattice model of glass},\ }\href
	{https://doi.org/10.1103/PhysRevB.95.184202} {\bibfield  {journal} {\bibinfo
			{journal} {Phys. Rev. B}\ }\textbf {\bibinfo {volume} {95}},\ \bibinfo
		{pages} {184202} (\bibinfo {year} {2017})}\BibitemShut {NoStop}%
	\bibitem [{\citenamefont {Lulli}\ \emph {et~al.}(2020)\citenamefont {Lulli},
		\citenamefont {Lee}, \citenamefont {Deng}, \citenamefont {Yip},\ and\
		\citenamefont {Lam}}]{lulli2020}%
	\BibitemOpen
	\bibfield  {author} {\bibinfo {author} {\bibfnamefont {M.}~\bibnamefont
			{Lulli}}, \bibinfo {author} {\bibfnamefont {C.-S.}\ \bibnamefont {Lee}},
		\bibinfo {author} {\bibfnamefont {H.-Y.}\ \bibnamefont {Deng}}, \bibinfo
		{author} {\bibfnamefont {C.-T.}\ \bibnamefont {Yip}},\ and\ \bibinfo {author}
		{\bibfnamefont {C.-H.}\ \bibnamefont {Lam}},\ }\bibfield  {title} {\bibinfo
		{title} {Spatial heterogeneities in structural temperature cause kovacs'
			expansion gap paradox in aging of glasses},\ }\href
	{https://doi.org/10.1103/PhysRevLett.124.095501} {\bibfield  {journal}
		{\bibinfo  {journal} {Phys. Rev. Lett.}\ }\textbf {\bibinfo {volume} {124}},\
		\bibinfo {pages} {095501} (\bibinfo {year} {2020})}\BibitemShut {NoStop}%
	\bibitem [{\citenamefont {Lee}\ \emph {et~al.}(2020)\citenamefont {Lee},
		\citenamefont {Lulli}, \citenamefont {Zhang}, \citenamefont {Deng},\ and\
		\citenamefont {Lam}}]{lee2020}%
	\BibitemOpen
	\bibfield  {author} {\bibinfo {author} {\bibfnamefont {C.-S.}\ \bibnamefont
			{Lee}}, \bibinfo {author} {\bibfnamefont {M.}~\bibnamefont {Lulli}}, \bibinfo
		{author} {\bibfnamefont {L.-H.}\ \bibnamefont {Zhang}}, \bibinfo {author}
		{\bibfnamefont {H.-Y.}\ \bibnamefont {Deng}},\ and\ \bibinfo {author}
		{\bibfnamefont {C.-H.}\ \bibnamefont {Lam}},\ }\bibfield  {title} {\bibinfo
		{title} {Fragile glasses associated with a dramatic drop of entropy under
			supercooling},\ }\href {https://doi.org/10.1103/PhysRevLett.125.265703}
	{\bibfield  {journal} {\bibinfo  {journal} {Phys. Rev. Lett.}\ }\textbf
		{\bibinfo {volume} {125}},\ \bibinfo {pages} {265703} (\bibinfo {year}
		{2020})}\BibitemShut {NoStop}%
	\bibitem [{\citenamefont {Lee}\ \emph {et~al.}(2021)\citenamefont {Lee},
		\citenamefont {Deng}, \citenamefont {Yip},\ and\ \citenamefont
		{Lam}}]{lee2021}%
	\BibitemOpen
	\bibfield  {author} {\bibinfo {author} {\bibfnamefont {C.-S.}\ \bibnamefont
			{Lee}}, \bibinfo {author} {\bibfnamefont {H.-Y.}\ \bibnamefont {Deng}},
		\bibinfo {author} {\bibfnamefont {C.-T.}\ \bibnamefont {Yip}},\ and\ \bibinfo
		{author} {\bibfnamefont {C.-H.}\ \bibnamefont {Lam}},\ }\bibfield  {title}
	{\bibinfo {title} {Large heat-capacity jump in cooling-heating of fragile
			glass from kinetic monte carlo simulations based on a two-state picture},\
	}\href {https://doi.org/10.1103/PhysRevE.104.024131} {\bibfield  {journal}
		{\bibinfo  {journal} {Phys. Rev. E}\ }\textbf {\bibinfo {volume} {104}},\
		\bibinfo {pages} {024131} (\bibinfo {year} {2021})}\BibitemShut {NoStop}%
	\bibitem [{\citenamefont {Gao}\ \emph {et~al.}(2022)\citenamefont {Gao},
		\citenamefont {Deng}, \citenamefont {Lee}, \citenamefont {You},\ and\
		\citenamefont {Lam}}]{gao2022}%
	\BibitemOpen
	\bibfield  {author} {\bibinfo {author} {\bibfnamefont {X.-Y.}\ \bibnamefont
			{Gao}}, \bibinfo {author} {\bibfnamefont {H.-Y.}\ \bibnamefont {Deng}},
		\bibinfo {author} {\bibfnamefont {C.-S.}\ \bibnamefont {Lee}}, \bibinfo
		{author} {\bibfnamefont {J.}~\bibnamefont {You}},\ and\ \bibinfo {author}
		{\bibfnamefont {C.-H.}\ \bibnamefont {Lam}},\ }\bibfield  {title} {\bibinfo
		{title} {Emergence of two-level systems in glass formers: a kinetic monte
			carlo study},\ }\href {https://doi.org/10.1039/d1sm01809d} {\bibfield
		{journal} {\bibinfo  {journal} {Soft Matter}\ }\textbf {\bibinfo {volume}
			{18}},\ \bibinfo {pages} {2211} (\bibinfo {year} {2022})}\BibitemShut
	{NoStop}%
	\bibitem [{\citenamefont {Gopinath}\ \emph {et~al.}(2022)\citenamefont
		{Gopinath}, \citenamefont {Lee}, \citenamefont {Gao}, \citenamefont {An},
		\citenamefont {Chan}, \citenamefont {Yip}, \citenamefont {Deng},\ and\
		\citenamefont {Lam}}]{gopinath2022}%
	\BibitemOpen
	\bibfield  {author} {\bibinfo {author} {\bibfnamefont {G.}~\bibnamefont
			{Gopinath}}, \bibinfo {author} {\bibfnamefont {C.-S.}\ \bibnamefont {Lee}},
		\bibinfo {author} {\bibfnamefont {X.-Y.}\ \bibnamefont {Gao}}, \bibinfo
		{author} {\bibfnamefont {X.-D.}\ \bibnamefont {An}}, \bibinfo {author}
		{\bibfnamefont {C.-H.}\ \bibnamefont {Chan}}, \bibinfo {author}
		{\bibfnamefont {C.-T.}\ \bibnamefont {Yip}}, \bibinfo {author} {\bibfnamefont
			{H.-Y.}\ \bibnamefont {Deng}},\ and\ \bibinfo {author} {\bibfnamefont
			{C.-H.}\ \bibnamefont {Lam}},\ }\bibfield  {title} {\bibinfo {title}
		{Diffusion-coefficient power laws and defect-driven glassy dynamics in swap
			acceleration},\ }\href {https://doi.org/10.1103/PhysRevLett.129.168002}
	{\bibfield  {journal} {\bibinfo  {journal} {Phys. Rev. Lett.}\ }\textbf
		{\bibinfo {volume} {129}},\ \bibinfo {pages} {168002} (\bibinfo {year}
		{2022})}\BibitemShut {NoStop}%
	\bibitem [{\citenamefont {Titantah}\ and\ \citenamefont
		{Karttunen}(2013)}]{titantah2013}%
	\BibitemOpen
	\bibfield  {author} {\bibinfo {author} {\bibfnamefont {J.~T.}\ \bibnamefont
			{Titantah}}\ and\ \bibinfo {author} {\bibfnamefont {M.}~\bibnamefont
			{Karttunen}},\ }\bibfield  {title} {\bibinfo {title} {Water dynamics:
			Relation between hydrogen bond bifurcations, molecular jumps, local density
			\& hydrophobicity},\ }\href@noop {} {\bibfield  {journal} {\bibinfo
			{journal} {Scientific reports}\ }\textbf {\bibinfo {volume} {3}},\ \bibinfo
		{pages} {1} (\bibinfo {year} {2013})}\BibitemShut {NoStop}%
	\bibitem [{\citenamefont {Debenedetti}(2003)}]{debenedetti2003}%
	\BibitemOpen
	\bibfield  {author} {\bibinfo {author} {\bibfnamefont {P.~G.}\ \bibnamefont
			{Debenedetti}},\ }\bibfield  {title} {\bibinfo {title} {Supercooled and
			glassy water},\ }\href@noop {} {\bibfield  {journal} {\bibinfo  {journal}
			{Journal of Physics: Condensed Matter}\ }\textbf {\bibinfo {volume} {15}},\
		\bibinfo {pages} {R1669} (\bibinfo {year} {2003})}\BibitemShut {NoStop}%
	\bibitem [{\citenamefont {Shi}\ \emph {et~al.}(2018)\citenamefont {Shi},
		\citenamefont {Russo},\ and\ \citenamefont {Tanaka}}]{shi2018}%
	\BibitemOpen
	\bibfield  {author} {\bibinfo {author} {\bibfnamefont {R.}~\bibnamefont
			{Shi}}, \bibinfo {author} {\bibfnamefont {J.}~\bibnamefont {Russo}},\ and\
		\bibinfo {author} {\bibfnamefont {H.}~\bibnamefont {Tanaka}},\ }\bibfield
	{title} {\bibinfo {title} {Origin of the emergent fragile-to-strong
			transition in supercooled water},\ }\href
	{https://doi.org/10.1073/pnas.1807821115} {\bibfield  {journal} {\bibinfo
			{journal} {Proc. Natl. Acad. Sci.}\ }\textbf {\bibinfo {volume} {115}},\
		\bibinfo {pages} {9444} (\bibinfo {year} {2018})},\ \Eprint
	{https://arxiv.org/abs/https://www.pnas.org/content/115/38/9444.full.pdf}
	{https://www.pnas.org/content/115/38/9444.full.pdf} \BibitemShut {NoStop}%
	\bibitem [{\citenamefont {Shi}\ and\ \citenamefont {Tanaka}(2020)}]{shi2020}%
	\BibitemOpen
	\bibfield  {author} {\bibinfo {author} {\bibfnamefont {R.}~\bibnamefont
			{Shi}}\ and\ \bibinfo {author} {\bibfnamefont {H.}~\bibnamefont {Tanaka}},\
	}\bibfield  {title} {\bibinfo {title} {The anomalies and criticality of
			liquid water},\ }\href@noop {} {\bibfield  {journal} {\bibinfo  {journal}
			{Proc. Natl. Acad. Sci.}\ }\textbf {\bibinfo {volume} {117}},\ \bibinfo
		{pages} {26591} (\bibinfo {year} {2020})}\BibitemShut {NoStop}%
	\bibitem [{\citenamefont {Moynihan}\ and\ \citenamefont
		{Angell}(2000)}]{moynihan2000}%
	\BibitemOpen
	\bibfield  {author} {\bibinfo {author} {\bibfnamefont {C.~T.}\ \bibnamefont
			{Moynihan}}\ and\ \bibinfo {author} {\bibfnamefont {C.~A.}\ \bibnamefont
			{Angell}},\ }\bibfield  {title} {\bibinfo {title} {Bond lattice or excitation
			model analysis of the configurational entropy of molecular liquids},\ }\href
	{https://doi.org/https://doi.org/10.1016/S0022-3093(00)00198-8} {\bibfield
		{journal} {\bibinfo  {journal} {J. Non-Cryst. Solids}\ }\textbf {\bibinfo
			{volume} {274}},\ \bibinfo {pages} {131} (\bibinfo {year}
		{2000})}\BibitemShut {NoStop}%
	\bibitem [{\citenamefont {Zhang}\ \emph {et~al.}(2021)\citenamefont {Zhang},
		\citenamefont {Wang}, \citenamefont {Yu},\ and\ \citenamefont
		{Douglas}}]{zhang2021}%
	\BibitemOpen
	\bibfield  {author} {\bibinfo {author} {\bibfnamefont {H.}~\bibnamefont
			{Zhang}}, \bibinfo {author} {\bibfnamefont {X.}~\bibnamefont {Wang}},
		\bibinfo {author} {\bibfnamefont {H.-B.}\ \bibnamefont {Yu}},\ and\ \bibinfo
		{author} {\bibfnamefont {J.~F.}\ \bibnamefont {Douglas}},\ }\bibfield
	{title} {\bibinfo {title} {Dynamic heterogeneity, cooperative motion, and
			johari--goldstein $\beta$-relaxation in a metallic glass-forming material
			exhibiting a fragile-to-strong transition},\ }\href@noop {} {\bibfield
		{journal} {\bibinfo  {journal} {The European Physical Journal E}\ }\textbf
		{\bibinfo {volume} {44}},\ \bibinfo {pages} {1} (\bibinfo {year}
		{2021})}\BibitemShut {NoStop}%
	\bibitem [{\citenamefont {Geske}\ \emph {et~al.}(2016)\citenamefont {Geske},
		\citenamefont {Drossel},\ and\ \citenamefont {Vogel}}]{geske2016}%
	\BibitemOpen
	\bibfield  {author} {\bibinfo {author} {\bibfnamefont {J.}~\bibnamefont
			{Geske}}, \bibinfo {author} {\bibfnamefont {B.}~\bibnamefont {Drossel}},\
		and\ \bibinfo {author} {\bibfnamefont {M.}~\bibnamefont {Vogel}},\ }\bibfield
	{title} {\bibinfo {title} {Fragile-to-strong transition in liquid silica},\
	}\href@noop {} {\bibfield  {journal} {\bibinfo  {journal} {AIP Advances}\
		}\textbf {\bibinfo {volume} {6}},\ \bibinfo {pages} {035131} (\bibinfo {year}
		{2016})}\BibitemShut {NoStop}%
	\bibitem [{\citenamefont {Zhai}\ \emph {et~al.}(2022)\citenamefont {Zhai},
		\citenamefont {Li}, \citenamefont {Wang}, \citenamefont {Hu}, \citenamefont
		{Song}, \citenamefont {Tian},\ and\ \citenamefont {Yue}}]{zhai2022}%
	\BibitemOpen
	\bibfield  {author} {\bibinfo {author} {\bibfnamefont {X.}~\bibnamefont
			{Zhai}}, \bibinfo {author} {\bibfnamefont {X.}~\bibnamefont {Li}}, \bibinfo
		{author} {\bibfnamefont {Z.}~\bibnamefont {Wang}}, \bibinfo {author}
		{\bibfnamefont {L.}~\bibnamefont {Hu}}, \bibinfo {author} {\bibfnamefont
			{K.}~\bibnamefont {Song}}, \bibinfo {author} {\bibfnamefont {Z.}~\bibnamefont
			{Tian}},\ and\ \bibinfo {author} {\bibfnamefont {Y.}~\bibnamefont {Yue}},\
	}\bibfield  {title} {\bibinfo {title} {The connection between the
			fragile-to-strong transition and the liquid-liquid transition in a binary
			alloy system},\ }\href@noop {} {\bibfield  {journal} {\bibinfo  {journal}
			{Acta Materialia}\ }\textbf {\bibinfo {volume} {239}},\ \bibinfo {pages}
		{118246} (\bibinfo {year} {2022})}\BibitemShut {NoStop}%
	\bibitem [{\citenamefont {Zhang}\ \emph {et~al.}(2010)\citenamefont {Zhang},
		\citenamefont {Hu}, \citenamefont {Yue},\ and\ \citenamefont
		{Mauro}}]{zhang2010}%
	\BibitemOpen
	\bibfield  {author} {\bibinfo {author} {\bibfnamefont {C.}~\bibnamefont
			{Zhang}}, \bibinfo {author} {\bibfnamefont {L.}~\bibnamefont {Hu}}, \bibinfo
		{author} {\bibfnamefont {Y.}~\bibnamefont {Yue}},\ and\ \bibinfo {author}
		{\bibfnamefont {J.~C.}\ \bibnamefont {Mauro}},\ }\bibfield  {title} {\bibinfo
		{title} {Fragile-to-strong transition in metallic glass-forming liquids},\
	}\href@noop {} {\bibfield  {journal} {\bibinfo  {journal} {J. Chem. Phys.}\
		}\textbf {\bibinfo {volume} {133}},\ \bibinfo {pages} {014508} (\bibinfo
		{year} {2010})}\BibitemShut {NoStop}%
	\bibitem [{\citenamefont {Bochtler}\ \emph {et~al.}(2017)\citenamefont
		{Bochtler}, \citenamefont {Gross},\ and\ \citenamefont
		{Busch}}]{bochtler2017}%
	\BibitemOpen
	\bibfield  {author} {\bibinfo {author} {\bibfnamefont {B.}~\bibnamefont
			{Bochtler}}, \bibinfo {author} {\bibfnamefont {O.}~\bibnamefont {Gross}},\
		and\ \bibinfo {author} {\bibfnamefont {R.}~\bibnamefont {Busch}},\ }\bibfield
	{title} {\bibinfo {title} {Indications for a fragile-to-strong transition in
			the high-and low-temperature viscosity of the fe43cr16mo16c15b10 bulk
			metallic glass-forming alloy},\ }\href@noop {} {\bibfield  {journal}
		{\bibinfo  {journal} {Applied Physics Letters}\ }\textbf {\bibinfo {volume}
			{111}},\ \bibinfo {pages} {261902} (\bibinfo {year} {2017})}\BibitemShut
	{NoStop}%
	\bibitem [{\citenamefont {Rosa~Jr}\ \emph {et~al.}(2020)\citenamefont
		{Rosa~Jr}, \citenamefont {Cruz}, \citenamefont {Santana}, \citenamefont
		{Brito},\ and\ \citenamefont {Moret}}]{rosa2020}%
	\BibitemOpen
	\bibfield  {author} {\bibinfo {author} {\bibfnamefont {A.~C.}\ \bibnamefont
			{Rosa~Jr}}, \bibinfo {author} {\bibfnamefont {C.}~\bibnamefont {Cruz}},
		\bibinfo {author} {\bibfnamefont {W.~S.}\ \bibnamefont {Santana}}, \bibinfo
		{author} {\bibfnamefont {E.}~\bibnamefont {Brito}},\ and\ \bibinfo {author}
		{\bibfnamefont {M.~A.}\ \bibnamefont {Moret}},\ }\bibfield  {title} {\bibinfo
		{title} {Non-arrhenius behavior and fragile-to-strong transition of
			glass-forming liquids},\ }\href@noop {} {\bibfield  {journal} {\bibinfo
			{journal} {Phys. Rev. E}\ }\textbf {\bibinfo {volume} {101}},\ \bibinfo
		{pages} {042131} (\bibinfo {year} {2020})}\BibitemShut {NoStop}%
\end{thebibliography}
\end{document}